\RequirePackage{fix-cm}
\documentclass{svjour3}                     
\smartqed  

\usepackage{graphicx}

\usepackage{float,bbm,natbib}
\usepackage{amsmath, amssymb, amsfonts}
\usepackage{xspace}
\usepackage{color,xcolor}

\renewcommand{\d}{\text{ d}}
\newcommand{\diag}{\text{diag}}
\newcommand{\logit}{\text{logit}}
\newcommand{\Esp}{\mathbb E}
\newcommand{\Bcal}{\mathcal{B}}

\newcommand{\Mcal}{\mathcal{M}}

\newcommand{\pt}{\widetilde{p}}

\newcommand{\pbar}{\overline{p}}
\newcommand{\papprox}{\pt_{\bY}}

\newcommand{\qapprox}{\pt_{\bY}}
\newcommand{\pexact}{p(\cdot | \bY)}
\newcommand{\ph}{p_h}
\newcommand{\rhoh}{\rho_h}

\newcommand{\Var}{\mathbb V}

\newcommand{\wtilde}{\widetilde{w}}

\newcommand{\gammabar}{\overline{\gamma}}

\newcommand{\balpha}{\boldsymbol{\alpha}}
\newcommand{\bbeta}{\boldsymbol{\beta}}

\newcommand{\bpi}{\boldsymbol{\pi}}
\newcommand{\bZ}{\boldsymbol{Z}}
\newcommand{\btheta}{\theta}

\newcommand{\bthetahm}{\btheta_h^m}
\newcommand{\bthetahmm}{\btheta_{h-1}^m}
\newcommand{\bthetah}{\btheta_h}
\newcommand{\bthetak}{\btheta_{k}}
\newcommand{\bthetakm}{\btheta_{k-1}}

\newcommand{\bthetaz}{\btheta_{0}}
\newcommand{\bthetazh}{\btheta_{0:h}}

\newcommand{\wh}{w_h}
\newcommand{\whm}{w_h^m}
\newcommand{\Whm}{W_h^m}
\newcommand{\Wh}{W_h}

\newcommand{\bY}{\boldsymbol{Y}}
\newcommand{\bx}{\boldsymbol{x}}
\newcommand{\bgamma}{\boldsymbol{\gamma}}

\newcommand{\SBMreg}{SBM-reg\xspace}
\newcommand{\SBS}{SBS\xspace}
\newcommand{\CBS}{CBS\xspace}
\newcommand{\CBSIS}{CBS$+$IS\xspace}

\newcommand{\vs}{\vspace{1em}}

\newtheorem{Remark}{Remark} 

\begin{document}

\title{Shortened Bridge Sampler: 
}
\subtitle{Using deterministic approximations to accelerate SMC for posterior sampling}


\author{Sophie Donnet         \and
        St\'ephane Robin 
}


\institute{UMR MIA-Paris, AgroParisTech, INRA, Universit\'e Paris-Saclay, 75005, Paris, France \\
  \email{[sophie.donnet][stephane.robin]@agroparistech.fr}}

\date{Received: date / Accepted: date}

\maketitle

\begin{abstract}
Sequential Monte Carlo has become a standard tool for Bayesian Inference of complex models. This approach can be computationally demanding, especially when initialized from the prior distribution. On the other hand, deterministic approximations of the posterior distribution are often available with no theoretical guaranties. We propose a bridge sampling scheme starting from such a deterministic approximation of the posterior distribution and targeting the true one. The resulting Shortened Bridge Sampler (SBS) relies on a sequence of distributions that is determined in an adaptive way. 

We illustrate the robustness and the efficiency of the methodology on a large simulation study. When applied to network datasets, SBS inference  leads to different statistical conclusions from the one supplied by the standard variational Bayes approximation.  

\end{abstract}

\section{Introduction \label{sec:Intro}}

In Bayesian statistics, except in a restricted number of conjugate models, the posterior distribution does not have a close form and requires the use of approximation methods. The 1990's have witnessed the developments of powerful stochastic methods and simulation-based algorithms able to perform Bayesian statistical inference on complex statistical models.

Among them, Monte Carlo Markov Chains (MCMC) --whose principle is to generate a Markov Chain such that its ergodic distribution is the posterior distribution (see for instance \cite{Robert:2005,RoC09} for an introduction)-- have been successfully applied to many problems, as attested by the countless publications in that domain. However, MCMCs suffer from several drawbacks. First of all, it is difficult to assess whether the chain has reached its ergodic distribution or not. Secondly, if the distribution of interest is highly multi-modal, MCMC algorithms can be trapped in local modes. More generally, when the space of parameters to be explored is of high dimension, MCMC algorithms will have difficulties in reaching their equilibrium distribution within a reasonable computational time. 

Recently, population based Monte Carlo methods have proved their efficiency and robustness in front of high dimensional and multimodal spaces. In a few words, population based Monte Carlo algorithms generate a large sample of parameters with a tractable distribution and update the importance sampling weights at each iteration, in order to finally match the distribution of interest. Among population based Monte Carlo, Sequential Monte Carlo (SMC) is a method combining parameters sampling and resampling. More precisely, a sequence of distributions of interest is designed, such that the first one is simple (i.e. easy to sample from) and the last one is the posterior distribution. This sequence of distributions defines the iterations of the algorithm. Then, at the first iteration, a sample of parameters is simulated with the first distribution. In the following iterations, the parameters are stochastically moved, weighted and resampled to follow the current distribution. The true posterior distribution is reached at the last iteration. The sequence of distributions can be dynamically designed. Primarily developed in the context of filtering problems (see \cite{Doucet2001}), they have been extended to the general problem of posterior sampling by \cite{DelMoral2006}. In comparison with MCMC methods, SMC does not require any burn-in period or convergence diagnostic. In addition, whereas computing marginal likelihood (for model comparison) has always been a challenging issue when using MCMC, SMC supplies an unbiased estimator of this quantity as a by-product of the algorithm. For all these reasons, SMC has proved its superiority over MCMC for complex models. 
 
In the recent years, particular fields (such as genomics or network analysis to name but a few) brought news statistical problems involving an increasing amount of data or statistical models with a large number of parameters. In such cases, not only MCMC but also population Monte Carlo algorithms have reached their limitations, requiring unreasonable computational time to explore the posterior distribution. To deal with such difficulties, deterministic approximations of the posterior distribution through optimization mathematical tools -- such as variational approximation (\cite{WaJ08, Blei2016}), Expectation-Propagation (\cite{Minka:2001}) or Integrated nested Laplace approximation (\cite{Rue09}) for instance-- have been proposed. These methods have the great advantage to be computationally light and can handle large data. However, their theoretical properties and accuracy is still under study. In particular, we do know that variational approximations can supply underestimated posterior variances (see for instance \cite{CoM07} for a large illustration of this phenomena on the Probit model).

On the one hand, SMC supplies a sample from the exact posterior distribution but can require unacceptable computational time. On the other hand, deterministic approximations (optimal in a sense to be determined) of the posterior distribution are fast methods but non-exact. 
One may therefore be tempted to take advantage of the two approaches in a combined strategy. The idea of combining variational Bayes inference with SMC is actually not new. 
\cite{RAJ15} split the data into block and compute the posterior distribution of $\theta$ given each block. They use a variational argument to propose the product of this partial posterior as a proxy for the true posterior. 
Focusing on Gaussian mixtures, \cite{McGPTAK16} consider online-inference and propose an sequential sampling scheme where, for each new batch of data, the variational approximation is iteratively updated and used as a proposal. 
\cite{Naesseth2017} use a SMC approach to get an improved, but still biased, variational approximation. 
Our approach is different from all these.
Our  main idea is to design a bridge sampling from the approximated posterior distribution to the true posterior distribution, the transfer from the approximate to the exact distribution being performed with an SMC algorithm (\cite{DelMoral2006}). The sampling method we propose can be considered from two points of view : either SMC is seen as a tool to correct the approximate distribution, or the approximate posterior distribution is seen as a mean to drastically accelerate the SMC procedure. 

Adopting the last perspective plunges the problematic into in a larger topic. Indeed, in general, for any challenging statistical model at stake, there exists a frequentist solution, suppling a point estimation of the parameter of interest. In the Bayesian practice, this point estimation is standardly set as an initial value in the MCMC algorithm, thus hoping a decrease of the computational time. In a SMC strategy starting with a sample from the prior distribution, such an initial point value is meaningless. We claim that the posterior distribution can be reached in a reduced computational time if the bridge sampling scheme starts from an approximate posterior distribution based on that point estimate.  We therefore refer to the proposed sampling method as the Shortened Bridge Sampler (\SBS).

\vspace{1em}

The paper is organized as follows. Section \ref{sec:Algo} is dedicated to the description of the methodology. 
We first remind the principle of importance sampling, then  introduce the sampling path and expose the algorithm in Subsection \ref{subsec:algo}. 
Its robustness and efficiency are illustrated on several simulated experiments in Section \ref{sec:Simul}. In Subsection \ref{sec:log}, the logit regression serves as a toy example to illustrate the computational time reduction and to test the robustness of the method with respect to the quality of the deterministic approximation of the posterior distribution. The Latent Class Analysis model (Subsection \ref{subsec:LCA}) is exploited to illustrate the relevance of our methodology on a mixture model, in particular, we propose a new strategy to tackle the label switching issue. On the Stochastic Block Model (SBM) with covariates (Subsection \ref{subsec:SBMreg}), we compare our strategy with the Variational Bayesian one in terms of model selection and model averaging. Finally, real datasets of social networks with covariates are presented in Section \ref{sec:Illust} : we stress the new insights brought by a ``correction'' of the Variational posterior approximation by the SMC strategy in term of both model averaging and significance of the parameters. Perspectives are discussed in Section \ref{sec:Discuss}. 


\section{From the approximate posterior distribution to the true posterior distribution} \label{sec:Algo}
Let us first introduce some notations.  $\bY$ denotes the observations, $\ell(\bY|\btheta)$ is the likelihood function with $\btheta \in \Theta$ the unknown parameters and $\pi(\btheta)$ is the prior distribution on $\btheta$. The Bayesian inference is based on the posterior distribution: 
\begin{equation}\label{eq:Bayes}
 p(\btheta | \bY) = \frac{\ell(\bY | \btheta) \pi(\btheta)}{p(\bY)}.
\end{equation}
where $p(\bY)$ is the marginal likelihood defined as: 
\begin{equation}\label{eq:marg}
 p(\bY) = \int \ell(\bY | \btheta) \pi(\btheta) d\btheta
\end{equation}
and is required in the Bayesian model selection procedure. 

\vspace{1em}

\noindent In what follows, $\papprox$ is an approximate posterior distribution on $\btheta$. \emph{We assume that $\papprox$ can be easily intensively simulated and that the density function of $\papprox$ has an explicit expression}  
The aim of this paper is to propose a way to use such an approximate posterior to actually sample from the true posterior.

\begin{Remark} Note that, in general, complex statistical models are written as hierarchical models and involve latent variables $\bZ$ (see Sections \ref{subsec:LCA} and \ref{subsec:SBMreg}). In such cases, the distributions of interest are the joint distribution $p(\bZ,\btheta | \bY)$ or the marginal one $p(\btheta | \bY)$. 
For the sake of simplicity, we chose to present the method without latent variables but obviously, all the following results and algorithms can be extended to this situation, replacing $\btheta$ by $(\bZ,\btheta)$. A substantial part of the simulations presented in Section \ref{sec:Simul} is devoted to such models.
\end{Remark}

\subsection{A first approach: Importance Sampling}\label{sec:is}
A first naive approach to use $\papprox$ consists in resorting to a simple importance sampling (IS) strategy, that is to say sampling $(\btheta^m)_{m = 1\dots M}$ from $\papprox$ and weighting the sample by $W^m \propto {\ell(\bY | \btheta^m) \pi(\btheta^m)}/{\papprox(\btheta^m)}$. However, this strategy is obviously naive for several reasons. First of all, there is no guarantee that the support of the approximate distribution includes the support of the true distribution, the contrary is even observed in practice for the variational approximation for instance \citep[see][]{CoM07,WaT05}. 
As a consequence, the posterior sample obtained through such an importance sampling strategy would be restricted to the support of $\papprox$ which can be strictly included in the support of $p(\cdot | \bY)$. Secondly, if $\papprox$ and $\pexact$ are very different, the sample will be degenerated, meaning that very few particles will have a non-negligible weight. This results in a small Effective Sample Size ($ESS$), which the algorithm we propose aims at keeping along iterations. In such situations, there is no hope to efficiently sample using 'one-step' IS but the principle can be used iteratively to progressively shift from the initial proposal to the true posterior distribution.

\subsection{A path sampling between the approximate and the true posterior distributions}

The main idea of this paper is to take advantage of the deterministic approximation  $\papprox$  of the posterior distribution to accelerate SMC procedure, or, inversely to transform sequentially a sample from the deterministic approximated posterior distribution into a sample from the true posterior distribution. 

\vspace{1em}

\noindent Sequential Monte Carlo samplers generate samples from a sequence of intermediate distributions $(\ph)_{h = 0\dots H}$ where the intermediate distributions $(\ph)_{h = 0\dots H}$ are smooth transitions from a simple distribution $p_0$ to the distribution of interest $p_H = p(\cdot | \bY)$. A classical choice for $(\ph)_{h = 0\dots H}$ \citep{Neal2001} is to consider: 
\begin{eqnarray}\label{eq:pih2}
 \ph(\btheta ) & \propto & \pi(\btheta) \ell(\bY | \btheta) ^{\rho_h}
\end{eqnarray}
where $\rho_0 = 0$, $\rho_H = 1$, thus slowly moving from the prior distribution to the posterior by progressively integrating the data $\bY$ through the likelihood function. In this paper, we propose an alternative scheme moving smoothly from the approximate posterior distribution $\papprox$ to the true $\pexact$. The path is thus defined by: 
\begin{eqnarray}\label{eq:pih}
 \ph(\btheta ) & \propto & \papprox(\btheta)^ {1-\rhoh}(p( \btheta | \bY)) ^{\rho_h}\nonumber\\
 & \propto & \papprox(\btheta)^ {1-\rhoh}( \pi(\btheta) \ell(\bY | \btheta)) ^{\rho_h}.
\end{eqnarray}
where,  $\rho_0 = 0$, $\rho_H = 1$. In a few words,  we start from the easy-to-sample distribution $ \papprox(\btheta)$ and progressively replace it with the true posterior distribution, this strategy being known as annealed importance sampling procedure \citep{Neal2001}. We claim that this scheme significantly reduces the computational time and is robust with respect to $ \papprox$. 

\begin{Remark}
Note that if $\papprox$ is chosen to be the prior distribution $\pi(\cdot)$, then schemes \eqref{eq:pih2} and \eqref{eq:pih} are identical. 
\end{Remark}

\begin{Remark}
An alternative strategy would consist in using $\papprox$ has an importance sampler in the first iteration of the standard annealing scheme defined in \eqref{eq:pih2}. However, in such a strategy, the approximate distribution $\papprox$ is under-exploited, since at the first iteration, the particles are reweighted with $W_0^m \propto {\pi(\btheta^m)}/{\papprox(\btheta^m |\bY)}$, thus going back, in practice, to a (possibly truncated) version of the prior distribution. This phenomena will be illustrated in Section \ref{sec:log}.
\end{Remark}

To sample from the sequence of distributions $(p_h)_{h = 1, \dots, H}$, we  adopt the sequential sampler proposed by \cite{DelMoral2006} where the annealing coefficients $\rho_h$ will be adjusted dynamically. We describe the algorithm in the following subsection. 

 \subsection{Shortened Bridge Sampling Algorithm}\label{subsec:algo}

 We now need to design an algorithm sequentially sampling from $ \ph(\btheta) \propto \papprox(\btheta)^ {1-\rhoh}( \pi(\btheta) \ell(\bY | \btheta)) ^{\rho_h}$. 
 The last years have witnessed a proliferation of scientific papers dealing with the problem of SMC methods and their applications \citep[see][ for an overview]{Doucet2001}. In our work, we resort to the algorithm proposed by \citet{DelMoral2006}. 
 
 \vs
 
\noindent Let us introduce the following notations: 
 \begin{eqnarray}\label{eq:gammah_Zh}
 \gamma_h(\theta) & = & \qapprox(\btheta)^ {1-\rho_h} \left[\ell(\bY | \btheta) \pi(\btheta)\right]^{\rho_h},
\end{eqnarray}
where 
$ Z_h = \int \gamma_h(\theta) \d \theta \nonumber $, 
so that $\ph(\theta) = \gamma_h(\theta) / Z_h$ is a probability density.
The main idea of \cite{DelMoral2006} is to plunge the problem of sampling a sequence of distributions defined on a single set $\Theta$ into the standard SMC filtering framework. To that purpose, the sequence $(\ph)_{h = 0\dots H}$ is replaced by a sequence of extended distributions:
 \begin{equation}\label{eq:pibarh}
\pbar_h(\bthetazh) = \frac{\gammabar_h(\bthetazh)}{Z_h}
\end{equation}
 with
 \begin{equation}\label{eq:pibarh2}
\gammabar_h( \bthetazh ) = \gamma_h(\bthetah) \prod_{k = 1}^{h} L_{k}\left(\bthetakm | \bthetak\right) 
\end{equation}
where $\bthetazh = (\bthetaz,\dots, \bthetah) \in \Theta \times \dots\times \Theta = \Theta^{h+1}$ and $(L_k)_{k = 0,\dots H-1}$ is a sequence of backward kernels satisfying: 
\begin{equation}\label{eq:L}
\int L_k\left(\bthetakm | \bthetak\right)d\btheta_{k-1} = 1, \quad \forall k = 0\dots H-1. 
\end{equation}
Due to Property \eqref{eq:L}, the marginal version of $\pbar_h$ (i.e. when integrating out $\bthetaz$ , $\dots$, $\btheta_{h-1}$) is the distribution of interest $\ph$. Once defined the sequence $(\pbar_h)_{h = 0\dots H}$, one may use the original SMC algorithm designed by \citet{Doucet2001} for filtering. At iteration $h$, the SMC sampler involves three steps: 
\begin{itemize}
\item \emph{Moving the particles} from $\btheta_{h-1}$ to $\btheta_{h}$ using a transition kernel $ K_h(\bthetah | \btheta_{h-1})$. As a consequence, let $ \eta_{h-1}(\btheta_{0:h-1})$ denote the sampling kernel for $ \btheta_{0:h-1}$ until iteration $h-1$, $\eta_h$'s expression is:
\begin{equation}\label{eq:eta}
 \eta_h(\bthetazh) = \eta_{h-1}(\btheta_{0:h-1}) K_h(\bthetah | \btheta_{h-1})
 \end{equation}
 \item \emph{Reweighing the particles} in order to correct the discrepancy between the sampling distribution $\eta_h$ and the distribution of interest at iteration $h$, $\pbar_h$. 
\item \emph{Selecting the particles} in order to reduce the variability of the importance sampling weights and avoid degeneracy. In practice the particles will be resampled when the  $ESS$  decreases below a pre-specified rate. 
\end{itemize}

\paragraph{About the importance weights.} 
 At iteration $h$, the importance sampling weights for $(\bthetazh^m)_{m = 1\dots M}$ are : $\forall m = 1\dots M$,
\begin{equation}\label{eq:w1}
w_h^m = \wh(\bthetazh^m) = \frac{\gammabar_h(\bthetazh^m)}{\eta_h(\bthetazh^m)}
\end{equation} 
in their unnormalized version. $(W^m_h)_{m = 1\dots M }$ denotes the normalized weights, i.e. 
\begin{equation}\label{eq:wW}
\Wh^m = \frac{\whm}{\sum_{m' = 1}^M w^{m'}_h}, \quad \forall m = 1\dots M
\end{equation}
Equations (\ref{eq:pibarh}-\ref{eq:pibarh2}-\ref{eq:eta}-\ref{eq:w1}) imply a recurrence formula for the weight of any particle $\bthetazh$: 
\begin{equation}\label{eq:w}
\wh(\bthetazh) = w_{h-1}(\btheta_{0:h-1}) \wtilde_{h-1:h}(\btheta_{h-1},\bthetah)
\end{equation}
where the incremental weight $ \wtilde_{h-1:h}(\btheta_{h-1},\bthetah)$ is equal to: 
 \begin{equation}\label{eq:wtilde}
 \wtilde_{h-1:h}(\btheta_{h-1},\bthetah) = \frac{L_h(\btheta_{h-1} | \btheta_h)}{ K_h(\bthetah | \btheta_{h-1})}\frac{\gamma_h(\bthetah)}{\gamma_{h-1}(\btheta_{h-1})}
\end{equation}

\paragraph{About the transition kernel $K_h$.} As, at this step, the target distribution is $\ph$, it seems natural to choose $K_h(\btheta_h | \btheta_{h-1})$ as a Monte Carlo Markov Chain (MCMC) kernel with $\ph(\btheta) \propto \qapprox(\btheta)^ {1-\rhoh}( \ell(\bY | \btheta) \pi(\btheta) ) ^{\rho_h}$ as stationary distribution. 
Following \cite{DelMoral2006}, we choose the backward kernel: 
\begin{equation}\label{eq:L2}
L_h(\btheta_{h-1} | \btheta_h) = \frac{ K_h(\btheta_h | \btheta_{h-1}) p_{h}(\btheta_{h-1}) }{\ph(\btheta_h)} 
\end{equation}
which satisfies Property \eqref{eq:L} and enables us to rewrite the weight increment $ \wtilde_{h-1:h}(\btheta_{h-1},\bthetah)$ appearing in \eqref{eq:w} and defined in \eqref{eq:wtilde} as
 \begin{equation}\label{eq:wtilde2}
 \wtilde_{h-1:h}(\btheta_{h-1},\btheta_h) = \frac{\gamma_{h}(\btheta_{h-1}) }{\gamma_{h-1}(\btheta_{h-1})} = \left[\alpha(\theta_{h-1})\right]^{\rho_h-\rho_{h-1}}
\end{equation}
where
\begin{equation}\label{eq:alpha}
\alpha(\theta) = \frac{\ell(\bY | \btheta) \pi(\btheta)}{\qapprox(\btheta| \bY)}. 
\end{equation}
In what follows, we denote $\alpha_h = \alpha(\btheta_h)$. 

\begin{Remark}
Using this particular backward kernel \eqref{eq:L2} has two major consequences. First it is not required having an explicit expression for the transition kernel $K_h(\btheta_h | \btheta_{h-1})$, which is quite welcome for MCMC kernels. Secondly, examining equations \eqref{eq:w} and \eqref{eq:wtilde2}, one may notice that the weight for a particle $\bthetazh$ does not depend on $\theta_h$ but only on $\btheta_{0:h-1}$. As a consequence, the weights of the particles $\bthetazh$ can be computed before they are simulated and for any new $\ph$
\end{Remark}

\paragraph{Adaptive design of $(\rho_h)_{h = 0\dots H}$.} 
As a consequence of this last remark, we are able to design an adaptive strategy for $(\rhoh)_{h = 0, \dots H}$ \cite[as in][]{Schafer2013,Jasra2011}. Indeed, being able to compute the weights of the up-coming particles for any new $\rho_h$, we can increase $\rho_h$ until the quality of the sample (measured through an indicator computed from the weights) decreases for the next distribution. In practice, 
following \cite{Zhou2016}, we use the conditional Effective Sampling Size ($cESS$) to measure the quality of $p_ {h-1}$ as an importance sampler when estimating an expectation against $\ph$. It is defined as: 
 \begin{eqnarray*}\label{eq:cESS}
 cESS & = & \left[\sum_{m = 1}^M M W_{h-1}^m \left(\frac{ \wtilde_{h-1:h}^m}{\sum_{m = 1}^M MW_{h-1}^m \wtilde_{h-1:h}^m} \right)^2 \right]^{-1} \\
 & = & \frac{M \left(\sum_{m = 1}^M W_{h-1}^m \wtilde_{h-1:h}^m\right)^2}{\sum_{m = 1}^M W_{h-1}^m ( \wtilde_{h-1:h}^m)^2}, 
\end{eqnarray*}
becoming 
\begin{eqnarray}\label{eq:cESS2}
&&cESS\left(\rho_h;\rho_{h-1}, (W_{h-1}^m,\alpha_{h-1}^m )_{m  \leq  M}\right) =cESS_{h-1}(\rho)\nonumber\\
&&
= \frac{M \left(\sum_{m = 1}^M W_{h-1}^m (\alpha_{h-1}^m)^{\rho_h -\rho_{h-1}}\right)^2}{\sum_{m = 1}^M W_{h-1}^m (\alpha_{h-1}^m)^{2(\rho_h -\rho_{h-1})}}. 
\end{eqnarray}
where $\alpha_{h-1}^m = \alpha(\btheta_{h-1}^m)$ has been defined in equation \eqref{eq:alpha}. 
If $\rho_h = \rho_{h-1}$ , $cESS$ si maximal (equal to $M$, the number of particles). As $\rho_h$ increases, the discrepancy between $p_ {h-1}$ and $\ph$ increases and so the quality of $p_ {h-1}$ as an importance sampling distribution when estimating an expectation against $\ph$ decreases and so does $cESS$. As a consequence, our strategy to find the next $\rho_h$ is to set: 
$$ 
 \rho_ h = 1 \wedge \sup_{\rho}\left\{\rho > \rho_{h-1}, cESS_{h-1}(\rho) \geq \tau_1 M \right\}
 $$

\paragraph{Selection of the particles.} In order to prevent a degeneration of the particle approximation, we use a standard resampling of the particles. In other words, if the variance of weights $(W_{h}^m)_{m = 1\dots M}$ is too high (or in other words, if the $ESS$ is too small), we resample the particles using a multinomial distribution, thus discarding the particles with low weights and duplicating the particles with high weights. 


\paragraph{Sampling algorithm.} 
Finally, we propose the following Shorten Bridge Sampling \SBS algorithm adapted to the sequence  (\ref{eq:pih}).

 \vspace{1em}
\noindent\rule{\columnwidth}{0.4pt}

\noindent \textbf{\SBS algorithm}
 \vspace{-0.5em}
 
\noindent\rule{\columnwidth}{0.4pt}

 \begin{enumerate}
 \item[] Set $(\tau_1,\tau_2) \in [0,1]^2$, $\rho_0 = 0$. 
 \item[0.] \emph{At iteration $0$} , sample $(\theta^m_{0})_{m = 1\dots M}$ from the approximate distribution $\papprox $. $ \forall m = 1\dots M$, set: 
 $$w_{0}^m = 1, \quad W_{0}^m = \frac{1}{M}, \quad \alpha_{0}^m = \frac{\ell(\bY | \btheta^m_{0}) \pi(\btheta^m_{0})}{\qapprox(\theta^m_{0})} $$
 \item[1.] \emph{At iteration $h$}: starting from $(\btheta_{h-1}^m, W_{h-1}^m,\alpha_{h-1}^m)_{m = 1\dots M}$
 \begin{enumerate}
 \item Find $\rho_h$ such that: 
 $$ 
 \rho_ h = 1 \wedge \sup_{\rho}\left\{\rho > \rho_{h-1},cESS_{h-1}(\rho) \geq \tau_1 M \right\},
 $$ 
 \item $\forall m = 1\dots M $, compute $\whm = w^m_{h-1}\left(\alpha_{h-1}^m\right)^{\rho_h - \rho_{h-1}}$ and $\Whm = {\whm}\left/{\sum_{m' = 1}^M w_h^{m'}}\right.$ 
 \item Compute 
 $$ESS_{h} = \frac{\left(\sum_{m = 1}^M\Whm\right)^2 }{\sum_{m = 1}^M (\Whm)^2} \in [1, M]$$ If $ESS_{h} < \tau_2\, M$, resample the particles 
 \begin{equation*}
 \begin{array}{ccl}
 (\btheta_{h-1}^m)' & \sim_{i.i.d} & \sum_{m = 1}^M \Whm \delta_{\{ \btheta^m_{h-1}\}}
 (\bthetahm)\\
 \bthetahmm & \leftarrow & (\btheta_{h-1}^m)'\\
 w_{h}^m &  \leftarrow  & 1 \\
 W_{h}^m & \leftarrow  & 1/M
 \end{array}
 \quad \forall m = 1\dots M
 \end{equation*}
 \item $\forall m = 1\dots M$, : propagate the particle $\bthetahm \sim K_h( \cdot | \btheta_{h-1}^m) $ where $K_h$ is a MCMC kernel with $\ph(\btheta) \propto \papprox(\btheta)^ {1-\rhoh}( \ell(\bY | \btheta) \pi(\btheta) ) ^{\rho_h}$ as an invariant distribution and compute:
 $$ \alpha^m_{h} = \frac{\ell(\bY | \btheta^m_{h}) \pi(\btheta^m_{h})}{\qapprox(\theta^m_{h}| \bY)} $$ 
 \end{enumerate}
 \item[2.] If $\rho_h = 1$, stop. If $\rho_h < 1$ return to $1$. 
\end{enumerate}
\vspace{-0.5em}
\noindent\rule{\columnwidth}{0.4pt}

\begin{Remark}
Let $\phi$ be a function defined in $\Theta$.  The study of the statistical properties of $\sum_{m=1}^ M W^H_m \phi(\theta^{H}_m)$  as an estimator of $\mathbf{E}[\phi(\theta)|\bY]$ is  a difficult task due to the sampling and resampling steps in the algorithm. However, many results can be found in the literature \citep[see][and references there in]{Doucet2009}.  First of all, $\sum_{m=1}^ M W^H_m \phi(\theta^{H}_m)$ is known to be strongly convergent. Moreover, following \cite{DelMoral2006}, a Central Limit Theorem can be obtained.  Besides, in addition to these asymptotic properties,  it is possible to control the mean-square error of  the estimator for a given  number of particles $M$, provided additional assumptions on $\phi$ .   Results of convergence were also provided by \cite{delmoral2012_conv} for adaptive sequential Monte Carlo algorithms. 

\end{Remark}

\subsection{Estimation of the marginal likelihood }

With respect to MCMC strategies, Annealing Importance Sampling and SMC have the great advantage to supply good estimators of the marginal likelihood. Indeed, as proved by \cite{DelMoral2006}, a non-biaised estimator of the marginal likelihood derives as a by-product of SMC. Moreover, the path sampling identity also provides an estimate of the marginal likelihood, as detailed hereafter.

\vs

\noindent Let us recall that $Z_ h = \int_{\Theta} \gamma_h(\btheta) d\theta$. Following \cite{DelMoral2006} and using the notations introduced below, the ratio of the marginal likelihoods ${Z_{h}} / {Z_{h-1}}$ is estimated by: 
\begin{equation*}
\widehat{\frac{Z_{h}}{Z_{h-1}}} = \sum_{m = 1}^M \Whm \wtilde_{h-1:h}^m. 
\end{equation*}
and 
\begin{equation}\label{eq:pZhat}
\widehat{\frac{Z_{H}}{Z_{0}}} = \prod_{h = 1}^H \widehat{\frac{Z_{h}}{Z_{h-1}}} = \prod_{h = 1}^H \sum_{m = 1}^M \Whm \wtilde_{h-1:h}^m 
\end{equation}
is an unbiased estimator of $Z_H/Z_0$. 
Now, $Z_H =  p(\bY)$ and $Z_0 = 1 $.

\noindent Another estimate is given by the path sampling identity. Indeed, under non-restrictive regularity assumptions, the following equality holds: 
\begin{equation}\label{eq:Int}
\log p(Y) - \log Z_0 = \int_{0}^{1} \mathbb{E}_{p_{\rho}}\left[\frac{d \log \gamma_{\rho}(\cdot)}{d \rho} \right]d\rho
\end{equation} 
where $\gamma_\rho(\theta) = \qapprox(\theta)^{1-\rho}(\ell(\bY | \theta) \pi(\theta))^{\rho}$, $p_\rho()$ is the associated probability density distribution and, in our geometric path sampling:
$$\frac{d \log \gamma_{\rho}(\cdot)}{d \rho} = \log \frac{\ell(\bY | \btheta) \pi(\btheta)}{\qapprox(\btheta)} = \log \alpha(\theta)$$
 An elementary trapezoidal scheme and Monte Carlo approximations of the expectations involved in \eqref{eq:Int} lead to the following approximation of the marginal likelihood:
\begin{equation}\label{eq:pZhat2}
\widehat{\widehat{\log p(\bY)}} = \sum_{h = 1}^H \frac{\rho_h - \rho_{h-1}}{2} (U^M_h + U^M_{h-1})
\end{equation}
where
 $U^M_h = \widehat{ \mathbb{E}}_{p_{\rho_h}}\left[ \log \alpha(\theta)\right] = \sum_{m = 1}^M W_{h}^m \log \alpha^m_{h}
 $. 

\begin{Remark}
Note that as suggested in \cite{Zhou2016}, we noticed on simulation studies that the two estimators behave similarly in our examples.  A precise comparison of the two estimators is out of the scope of this paper. 
\end{Remark}

\section{Simulation study \label{sec:Simul}}

We now present a large simulation study. The goal of the study is to  assess the fact that the proposed \SBS algorithm --that combines  an optimization-based approximation of the posterior distribution with a SMC sampler-- drastically decreases the computational time with respect to a classical annealing-scheme or, equivalently, that the approximated posterior distribution can be corrected into the true posterior  distribution at a low computational cost. 

\subsection{Logistic regression}\label{sec:log}
 This first model is used as a toy example to illustrate the efficiency and the robustness of our methodology. 
Let $(Y_1, \dots, Y_n)$ be a set a $n$ independent observations with values in $\{0,1\}$. Any individual observation $i$ is described by a vector $\bx_i \in \mathbb{R}^p$ of $p$ covariates and we consider the logistic regression model: 
$$ P(Y_i = 1) = 1-P(Y_i = 0) = \frac{e^{\bx_i^t \theta}}{1+e^{\bx_i^t \theta} }$$

 \noindent We generate a simulated dataset with a randomly chosen regression matrix $X$ and the following parameters: 
$$ n = 200, \quad p = 4, \quad \theta = (0.5, -0.6, 0, -1).$$

\begin{figure}
\centering
\includegraphics[width = 0.45\textwidth]{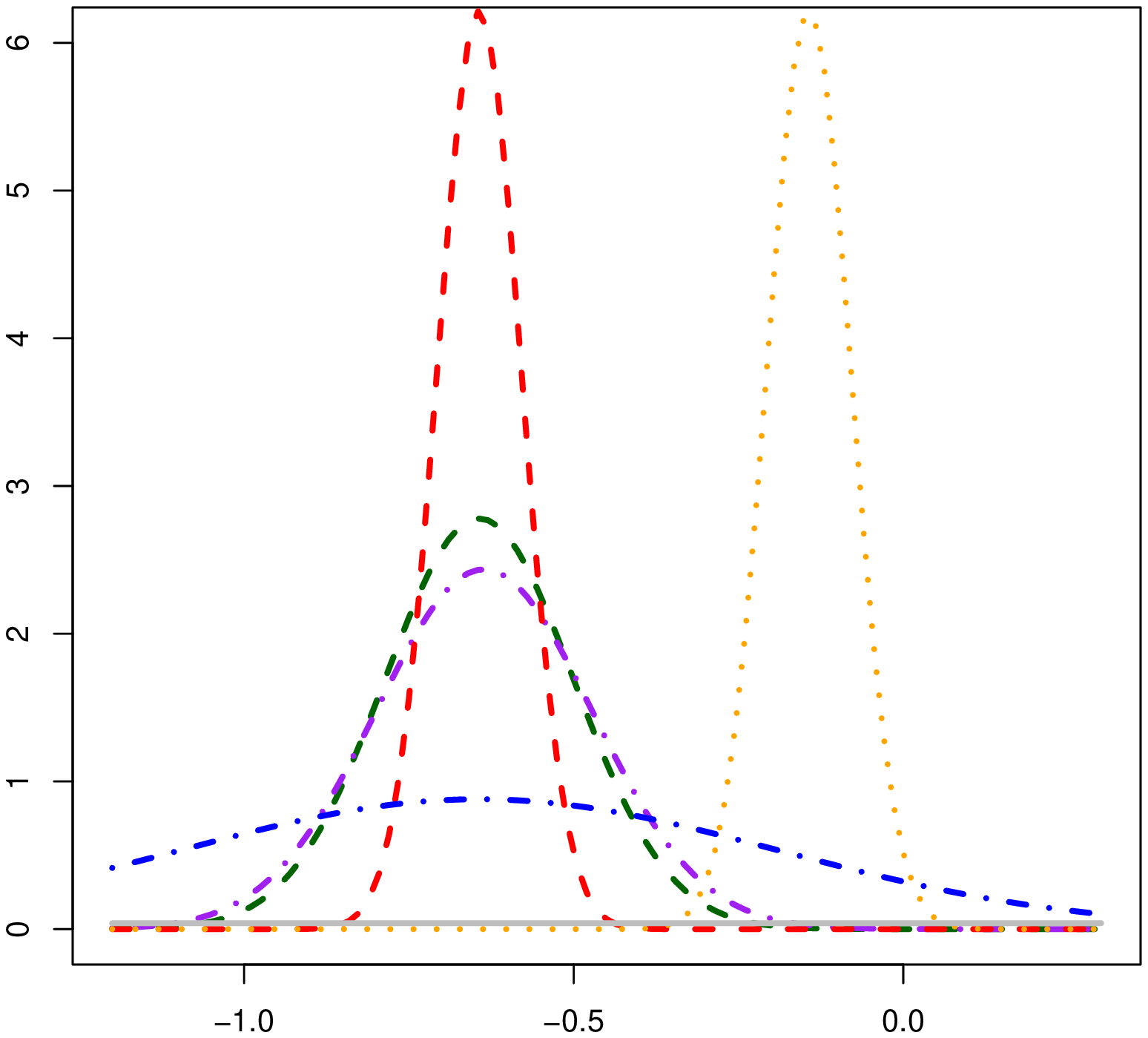}
\includegraphics[width = 0.45\textwidth]{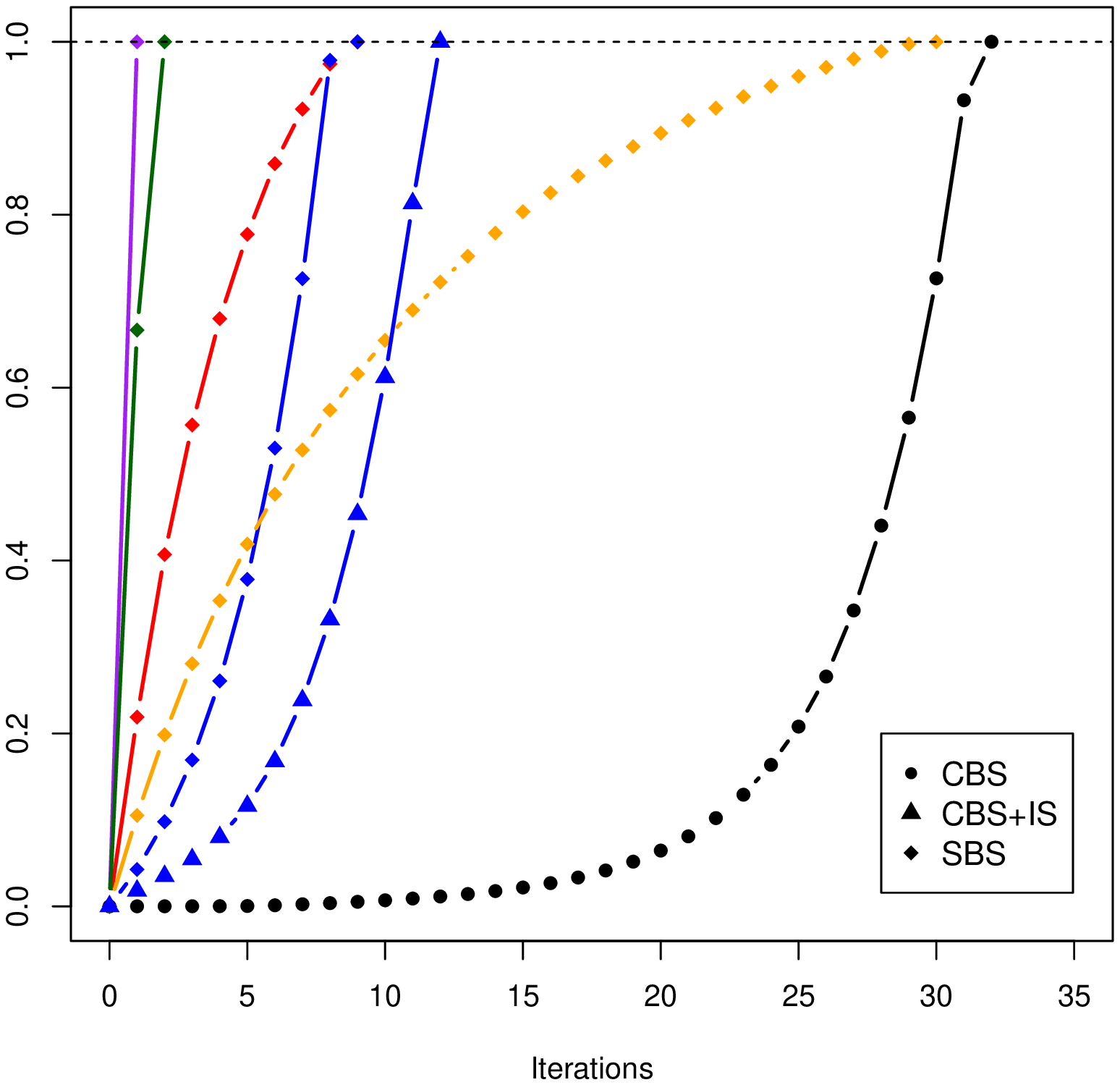}

\caption{\textbf{Logistic regression}. \emph{Left frame} :  approximate posterior distributions for $\theta_2$. $\widetilde{p}_{\bY,1}$ (VB) is in the darkgreen curve, $\widetilde{p}_{\bY,2}$ (ML) is in purple, $\widetilde{p}_{\bY,3}$ (VB with decreased variance) is in red, $\widetilde{p}_{\bY,4}$ (VB with increased variance) is in blue and $\widetilde{p}_{\bY,4}$ (VB shifted with small variance) is in orange. \emph{Right frame}: sequences $(\rho_h)_{h\leq 0}$ with the three algorithms and the $5$ $\papprox$ along the iterations of algorithm. The result for \CBS is the dotted black line. The results for our algorithm \SBS are labeled by diamonds $\Diamond$ whereas the ones for algorithm \CBSIS are marked by triangles $\triangle$.}
\label{fig:logreg:approx}
\end{figure}

\noindent Setting a Gaussian prior distribution on $\theta \in \mathbb{R}^p$, $\theta \sim \mathcal{N}(0,\sigma^2\boldsymbol{I}_p),$ with $\sigma^2 = 100$, it is natural to propose a Gaussian approximation of the posterior distribution $\papprox : = \mathcal{N}(\widehat{\mu},\widehat{\Sigma})$. We first  computed  the Gaussian variational Bayes estimator and we obtained:
 $$\widehat{\mu}^{VB} = (0.398, -0.643, -0.280, -0.847) $$ and 
$$ \widehat{\Sigma}^{VB} = 
 10^{-3} \left(
 \begin{array}{rrrr}
 23.3 & -1.2 & 0.1 & 1.1 \\ 
 -1.2 & 20.6 & -0.1 & 2.0 \\ 
 0.1 & -0.1 & 24.0 & 1.4 \\ 
 1.1 & 2.0 & 1.4 & 25.5 
 \end{array}
 \right)
 . 
 $$
We also  considered an approximate Gaussian posterior distribution based on the maximum likelihood estimator (ML) $\widehat{\theta}^{ML}$. The ML and its asymptotic variance $ \widehat{\Sigma}^{ML}$ are obtained with the \textsf{R}-function \textsf{glm}. 
To test the robustness of our bridge sampling with respect to $\papprox$, we also consider an artificially increased (respectively decreased) variance. In addition, we consider a distribution centered on an aberrant value with a small variance, thus leading to five different $\papprox$: 
\begin{eqnarray*}
\widetilde{p}_{\bY,1} & = & \mathcal{N}(\widehat{\mu}^{VB}, \widehat{\Sigma}^{VB})\\
\widetilde{p}_{\bY,2} & = & \mathcal{N}(\widehat{\theta}^{ML}, \widehat{\Sigma}^{ML})\\
 \widetilde{p}_{\bY,3} & = & \mathcal{N}(\widehat{\mu}^{VB}, \diag(\widehat{\Sigma}^{VB} )/5)\\
 \widetilde{p}_{\bY,4} & = & \mathcal{N}(\widehat{\mu}^{VB}, \diag(\widehat{\Sigma}^{VB} )\times10)\\
 \widetilde{p}_{\bY,5} & = & \mathcal{N}(\widehat{\mu}^{VB} + 0.5, \diag(\widehat{\Sigma}^{VB} )/5)
 \end{eqnarray*}
$\widetilde{p}_{\bY,1}$, $\widetilde{p}_{\bY,2}$, $\widetilde{p}_{\bY,3}$ $\widetilde{p}_{\bY,4}$ and $\widetilde{p}_{\bY,5}$ are plotted in Figure \ref{fig:logreg:approx} (for $\theta_2$). 

\vspace{1em}

\noindent For each $(\widetilde{p}_{\bY,k})_{k = 1\dots 5}$, we sample the posterior distribution using three methods. 

\begin{itemize}
\item \sloppy \CBS refers to a Classical Bridge Sampling $\pi(\theta)\ell(\bY| \theta)^ {\rho_h}$, sequentially sampled with a SMC algorithm. The strategy serves as a reference to be compared to the other ones. 
\item \CBSIS refers to the same annealing scheme as \CBS  but the first sample $(\theta_0^m)_{m=1,\dots,M}$ is generated with  $\widetilde{p}_{\bY}$ and the adequate weights are computed. 
\item \sloppy Finally, we use \SBS, described in section \ref{subsec:algo} corresponding to the annealing scheme:   $ \papprox(\btheta)^ {1-\rhoh}( \pi(\btheta) \ell(\bY | \btheta)) ^{\rho_h}$.
\end{itemize}

\paragraph{Tunings.}
In each case, the SMC is performed using $M = 10000$ particles, $\tau_1 = 0.9$ and $\tau_2 = 0.8$. The kernel $K_h$ is a composed of $B = 5$ iteration of a standard Metropolis-Hastings kernel, proposing a new parameter as: 
 $ \theta^c \sim \frac{1}{3} \sum_{i = 1}^3 \mathcal{N}(\theta^{\ell-1}, \rho_i \times \widehat{\Sigma}^{ML})$ 
 where $\widehat{\Sigma}^{ML}$ is the asymptotic variance of the maximum likelihood estimator and $(\rho_1,\rho_2,\rho_3) = (1,0.1,10)$.

\begin{figure} 
\centering

\includegraphics[width = 0.45 \textwidth]{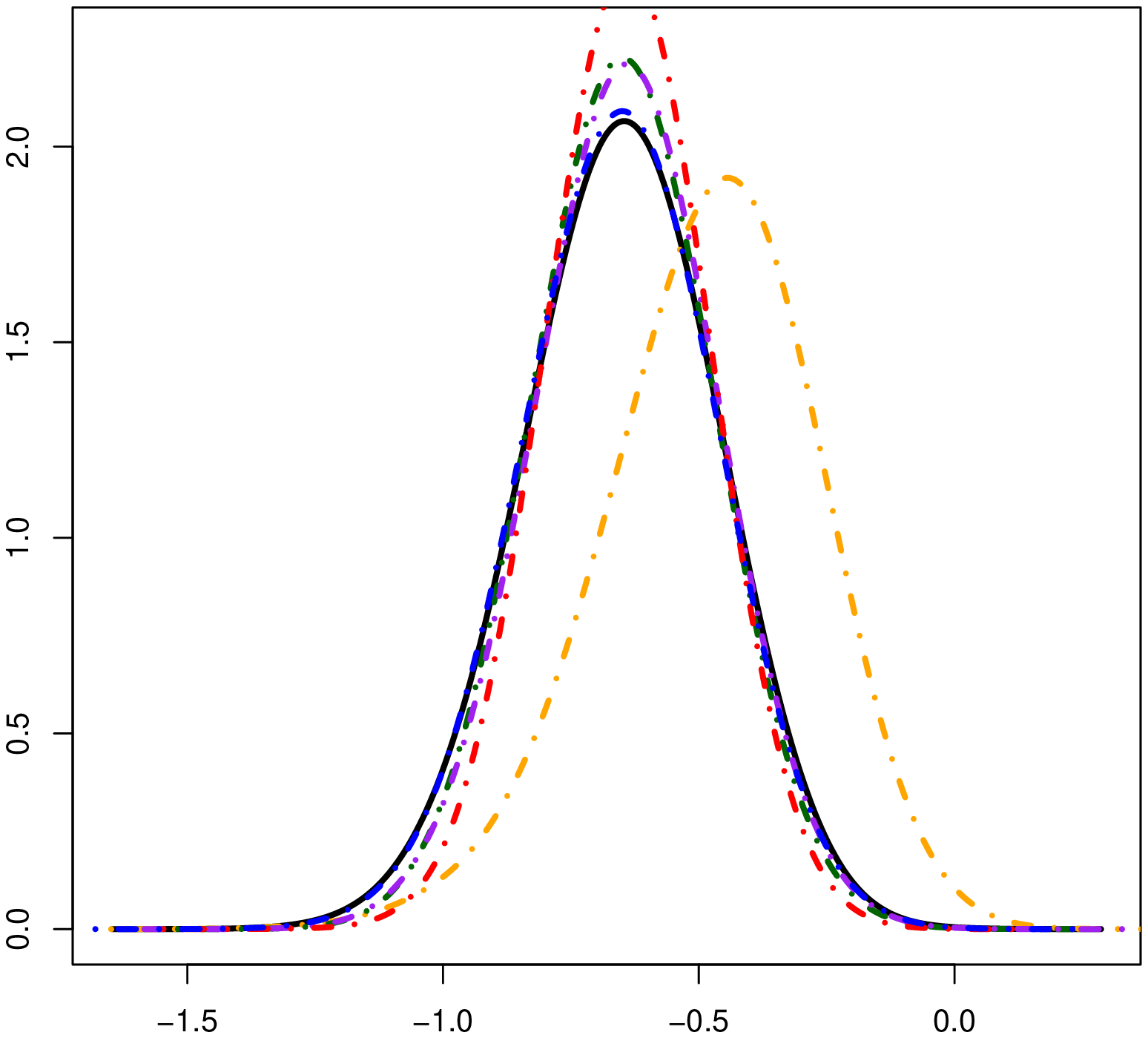}
\includegraphics[width = 0.45\textwidth]{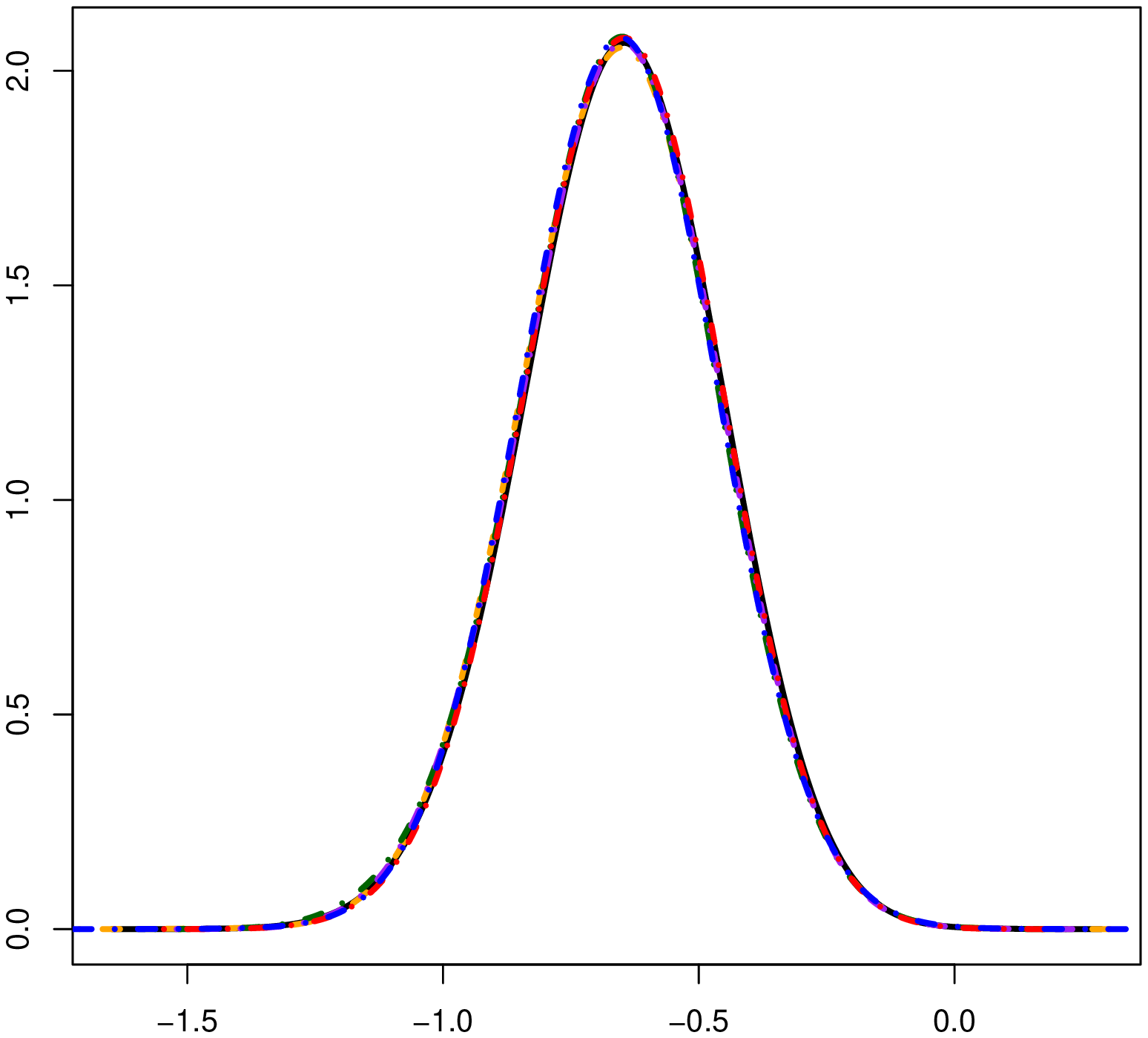}

 \caption{\textbf{Logistic regression}. Posterior distributions for $\theta_2$ obtained with algorithm \CBSIS (left) and our strategy \SBS (right) with the $5$ $(\widetilde{p}_{\bY,k})_{k = 1,\dots,5}$ with the following colors: $\widetilde{p}_{\bY,1}$ (VB) in darkgreen. $\widetilde{p}_{\bY,2}$ (ML) in purple. $\widetilde{p}_{\bY,3}$ (VB with decreased variance) in red. $\widetilde{p}_{\bY,4}$ (VB with increased variance) in blue. $\widetilde{p}_{\bY,4}$ (VB translated with small variance) in orange. The results for \CBS are plotted in black. 
 }. 
\label{fig: log reg post}
\end{figure}

\paragraph{Results.} The results are plotted in Figures \ref{fig: log reg post} and \ref{fig:logreg:approx} (right). For the five $(\widetilde{p}_{\bY,k})_{k = 1 \dots 5}$, the posterior distributions given by our algorithm \SBS (Figure \ref{fig: log reg post}, right) are confounded and stick to the one obtained from the reference algorithm \CBS, thus illustrating the practical robustness of our new bridge sampler even. More precisely, even if the approximated posterior distribution has an underestimated variance (which is known to be the case for the Variational Bayes estimator in this case), our methodology will supply a sample from the true posterior distribution. Moreover, the algorithm is also robust when $\papprox$ is piked around on an absurd value (see results for $\widetilde{p}_{\bY,5})$). 

On the contrary, using algorithm \CBSIS can be a bad idea. Indeed, if the approximation $\widetilde{p}_{\bY}$ has a wider support than the true posterior, the algorithm will perform well and provide a sample from the right posterior distribution: as can be seen in Figure \ref{fig: log reg post}, left frame, the black and the blue curves are indistinguishable. However, if $\widetilde{p}_{\bY}$ has an underestimated variance ($\widetilde{p}_{\bY,1}$, $\widetilde{p}_{\bY,2}$, $\widetilde{p}_{\bY,4}$ and $\widetilde{p}_{\bY,5}$ ), the standard sampling strategy will lead to a false posterior distribution (see the orange, red, purple and green curves in the left frame of Figure \ref{fig: log reg post}). Note that, there is no algorithmic indicator detecting such a bad behavior.

\noindent To compare the computational times of the various strategies, we can have a look at the number of iterations required for the sequences $(\rho_h)_{h \geq 0}$ to reach $1$. These sequences $(\rho_h)_{h \geq 0}$ are plotted in Figure \ref{fig:logreg:approx} (right). We only plot the curves for the combinations ``algorithm/approximated posterior $\widetilde{p}_{\bY}$'' leading to the true posterior distribution. As expected, \CBS is the most time consuming (see black dotted curve). As an indicative basis, on this example, the $30$ iterations of  \CBS  requires roughly $5$ minutes on an  \textsf{Intel \textregistered Xeon(R) CPU E5-1650 v3 @ 3.50GHz $\times$ 12}  using  six  cores. With $\widetilde{p}_{\bY,3}$ (increased variance), \SBS and \CBSIS finish in quite comparable computational time with slightly better results for our methodology. 
\SBS requires the same computational time with $\widetilde{p}_{\bY,4}$ (small variance). 
 \SBS clearly outperforms \CBS. Finally, considering an extreme case where the approximation distribution $\widetilde{p}_{\bY,5}$ is concentrated around an aberrant value (which is unlikely to be the case in practice), the \SBS and \CBS have comparable computation times. 
 
 \vspace{1em}

\noindent As a conclusion, \CBSIS can not be used is general cases since it can supply a wrong approximation of the posterior distribution. This is due to the fact that, at the first iteration of \CBSIS, the sample must be of the prior distribution. Using IS, i.e. simulating with $\widetilde{p}_{\bY,i}$ and assessing weight can lead to a sample from the truncated prior distribution $\pi(\btheta)\mathbbm{1}_{\widetilde{p}_{\bY}(\btheta)>0}$. This simulation from the wrong distribution at the first step of the sequential importance sampler is not corrected in the following iterations. 
Besides, we are not able to detect such a phenomena. 

\noindent On the contrary, our new bridge sampler behaves well, whatever $\papprox$. An under-evaluated variance in $\papprox$ is not a limit to the use of our scheme. The gain in computational time depends obviously on the distance between $\papprox$ and $p(\cdot | \bY)$ but is expected to be drastic when $\papprox$ comes from deterministic approximations of the posterior (Variational Bayes, Expectation Propagation, etc).


\subsection{Latent Class Analysis model} \label{subsec:LCA}
In this section we consider the latent class analysis (LCA) model. On this model, we focus on the label switching issue and show that the strategy we propose can tackle this difficulty.

\subsubsection{Model and prior distribution}
LCA is a mixture model for multivariate binary observations such as the correct or incorrect answers submitted during an exam \citep{Barthol11}, the symptoms presented by persons with major depressive disorder \citep{Garrett00} or a disability index recorded by long-term survey \citep{erosheva2007}. 
Let $(\bY_i)_{i = 1,\dots, n} = (Y_{i1},\dots, Y_{iq})_{i = 1,\dots, n}$ be $n$ i.i.d observations where $\forall (i,j)$, $Y_{ij} \in \{0,1\}$. $i$ and $j$ are respectively the individual and the response indices. Each $\bY_i$ is assumed to arise from the following mixture model: 
$$
\mathbb{P}(\bY_i = (y_{i1},\dots, y_{iq})) = \sum_{k = 1}^g \pi_k \prod_{j = 1}^q \gamma_{kj}^{y_{ij}} (1-\gamma_{kj})^{1-y_{ij}}
$$
where $\pi_k$ represents the proportion of the $k$-th component ($\sum_{k = 1}^g \pi_k = 1$) and $\gamma_{kj}$ is the success probability for the $j$-th response in the $k$-th group. 
The model is equivalently written as
\begin{equation}
\begin{array}{cclr}
Y_{ij} | Z_i & \sim & \mathcal{B}(\gamma_{Z_i \, j}), & \quad \quad \forall i = 1,\dots n, j = 1,\dots, q\\
\mathbb{P}(Z_{i} = k) & = & \pi_k, & \quad \quad \forall i = 1,\dots n, k = 1,\dots, g
\end{array}
\end{equation}
where $\bZ = (Z_1,\dots, Z_n)$ is a latent random vector. 

\vspace{1em}

\noindent \sloppy We set the following standard exchangeable prior distributions on $\bpi = (\pi_1,\dots, \pi_g)$ and $\bgamma = (\gamma_{kj})_{k = 1,\dots,q, j = 1,\dots,q}$:
\begin{equation} \label{eq:LCAprior}
\begin{array}{cccl}
(\pi_1,\dots, \pi_g) & \sim & & \mathcal{D}ir(d, \dots d)\\
(\gamma_{kj})_{k = 1,\dots,g, j = 1,\dots,q} & \sim & _{i.i.d} & \mathcal{B}eta(a,b) 
\end{array}
\end{equation}

\subsubsection{Posterior distribution and label switching} 
As for any mixture model, the posterior distribution should reproduce the likelihood invariance under permutation of the mixture indices. In other words, the posterior distribution is multi-modal, each mode corresponding to a permutation of the index of the mixture components. In such cases, it is well documented that MCMC algorithms often fail into exploring the various modes of the posterior distribution. Note that the label switching issue arises not only when it comes out to sample the posterior distribution but also for evidence approximation in model selection \citep[see for instance the introduction of ][and references inside]{Lee2016}. In this section, we illustrate the fact that a simple solution --based on the Variational Bayes posterior approximation and our sampling algorithm \SBS-- can be proposed to handle the label switching problem.

\vspace{1em}

\noindent
On this model, it is easy to derive a mean-field variational approximation of the posterior distribution, resulting into a posterior approximation of the form: 
\begin{eqnarray}\label{eq:LCApostVB}
 &&\papprox^{\mathit{VB}}(\bZ,\bgamma, \bpi) =f_{ \mathcal{D}ir(\tilde \delta_1, \dots \tilde \delta_g)}(\bpi)\nonumber \\
 && \times \prod_{k = 1,j=1}^{g,q}   f_{ \mathcal{B}eta(\tilde \alpha_{kj},\tilde \beta_{kj})} (\gamma_{kj})   \prod_{i = 1}^n \prod_{k = 1}^ q (\tau_{ik})^{\mathbbm{1}_{Z_{i} = k}} 
\end{eqnarray}
Details can be found in \cite{White13} and the algorithm is implemented in the corresponding R-package \textsf{BayesLCA}. 
However, contrary to the true posterior distribution $(p(\cdot | \bY)$, $\papprox^{\mathit{VB}}$ is not exchangeable. 
 Moreover, this posterior approximation is known to be excessively concentrated around one mode. As a consequence, we can presume (and we illustrate it on the following numerical experiments) that our sampling algorithm \SBS starting from $ \papprox$ and using a standard Gibbs transition kernel will not be able to propagate particles on the other modes of the posterior distribution. 
\begin{Remark}
When talking about a \emph{``standard Gibbs transition kernel''}, we refer to the most naive Gibbs algorithm, sequentially sampling $[\bZ | \bY, \bgamma, \bpi]$, $[\bgamma | \bY, \bZ \bpi]$ and $[\bpi | \bZ]$. Using the expressions for $ \papprox^{\mathit{VB}}$ and $p_{\rho_n}(\bZ,\bgamma,\bpi)$, these three distributions are conjugate. We stick to this MCMC kernel, and do not introduce any modification to \emph{force or prevent} the label switching phenomena at the propagation step. 
\end{Remark}

\noindent $ \papprox^{\mathit{VB}}$ being unsatisfying from the label switching perspective, we introduce its so called \emph{symmetrized version}, forcing the invariance by permutation. More precisely, let $\mathcal{S}_g$ be the set of all the permutations of $\{1,\dots,g\}$, we define 
\begin{equation} \label{eq:LCApostVBsym0}
\papprox^{\mathit{VB.Sym}}(\bZ,\bgamma, \bpi) = \sum_{\sigma \in \mathcal{S}_q} \papprox^{\mathit{VB.Sym}}(\bZ,\bgamma, \bpi, \sigma)
\end{equation}
where 
\begin{eqnarray}\label{eq:LCApostVBsym}
 &&\papprox^{\mathit{VB.Sym}}(\bZ,\bgamma, \bpi, \sigma) 
  = \frac{1}{g !} f_{ \mathcal{D}ir(\tilde \delta_{\sigma(1)}, \dots \tilde \delta_{\sigma(g)})}(\bpi)
 \nonumber\\
 & & \times \prod_{k = 1}^g \prod_{j = 1}^ q f_{ \mathcal{B}eta(\tilde \alpha_{\sigma(k)j},\tilde \beta_{\sigma(k)j})} (\gamma_{kj}) 
 \prod_{i = 1}^n \prod_{k = 1}^ q (\tau_{i\sigma(k)})^{\mathbbm{1}_{Z_{i} = k}}.\nonumber\\ 
\end{eqnarray}

\noindent In order to keep the conjugacy properties in conditional distributions we use our algorithm \SBS to sequentially sample: 
 \begin{eqnarray}\label{eq:LCApostVBsym2}
 p_{\rho_h}(\bZ,\bgamma, \bpi, \sigma ) \propto [\papprox^{\mathit{VB.Sym}}(\bZ,\bgamma, \bpi, \sigma)]^ {1-\rho_h} && \nonumber  \\
  \times [ \ell(\bY | \bZ, \bgamma,\bpi)p(\bZ | \bpi) \pi(\bpi,\bgamma)]^ {\rho_h}&&
 \end{eqnarray}
Once $\sigma$ has been integrated out, our sequential sampling scheme starts at $(\papprox^{\mathit{VB.Sym}})$ given in equation \eqref{eq:LCApostVBsym0} and terminates at the true posterior distribution.

\subsubsection{Simulation design and test}\label{subsubsec:criteria}

This example aims at proving that our strategy supplies a sample from the true posterior distribution, even in the difficult framework of mixture models. To assess the validity of our method, we propose using the following testing procedure. 

\paragraph{Testing procedure.} We introduce the following property from which we derive a negative criterion, in the sense that if the obtained distribution is the true posterior distribution, it must satisfy the criterion.

\begin{property} \label{lem:criterion}
\noindent Let $\Phi: \Theta \mapsto \mathbb{R}$ be such that $\exists \; \Psi$ verifying $ H: \btheta \mapsto \left(\Phi(\btheta), \Psi(\btheta)\right)$ is injective and continuously differentiable.Assume that 
$$ (\btheta^\star, \bY^{\star}) \sim \pi(\btheta^\star)\ell(\bY^\star | \btheta ^\star). $$ 
Given $\bY^\star$, let $p_{\bY^\star}(\Phi(\theta))$ be any probability distribution on $\Phi(\Theta)$. Let $U(\btheta^\star, \bY^\star,\Phi,p_{\bY^\star})$ be the following statistic: 
\begin{equation} \label{eq:crit-theo}
 U(\btheta^\star, \bY^\star,\Phi,p_{\bY^\star}) = \mathbb{E}_{p_{\bY ^\star}} \left[ \mathbbm{1}_{\Phi(\btheta) < \Phi(\btheta ^{\star})}\right].
\end{equation}
\noindent Then, if $
p_{\bY^\star}(\Phi(\theta)) = p(\Phi(\btheta)|\bY^\star)$,  $  \forall \btheta \in \Theta$ then 
$U(\btheta^\star, \bY^\star,\Phi,q_{\bY^\star}) \sim \mathcal{U}_{[0,1]}$.  
\end{property}

\begin{Remark}
Note that when $U(\btheta^\star, \bY^\star,\Phi, p_{\bY^\star})$ has no explicit expression and when we have access to a sample from $q_{\bY^\star}$, we can replace $U(\btheta^{\star}, \bY^{\star },\Phi, p_{\bY^{\star}})$ by its non-biased and convergent estimator: 
\begin{equation} \label{eq:crit-emp}
U_M(\btheta^{\star},\bY^{\star},\Phi, (\btheta_m)_{m = 1\dots M}) = \frac{1}{M}\sum_{m = 1}^M \mathbbm{1}_{\Phi(\theta_m) <\Phi(\btheta^{\star}) }
\end{equation}
where $\btheta_m \sim_{i.i.d.} p_{\bY^\star}$
and moreover if $p_{\bY^\star}(\Phi(\theta)) = p(\Phi(\btheta)|\bY^\star)$ for all $\btheta \in \Theta$, then
$$
U_M(\btheta^\star,\bY^\star,\Phi, (\btheta_m)_{m = 1\dots M}) \sim \mathcal{U}_{\left\{0,\frac{1}{M},
\frac{2}{M},\dots, 1\right\}}.
$$ 
\end{Remark}

\vspace{1em}

\noindent This property enables the elaboration of our checking procedure in $4$ steps.

\vspace{1em}
\noindent\rule{\columnwidth}{0.4pt}

\noindent \textbf{Checking procedure for posterior approximation} 

\vspace{-0.5em}
\noindent\rule{\columnwidth}{0.4pt}

\noindent For a given approximation method $\Mcal$ of the posterior distribution. Case (a) corresponds to deterministic approximations and case (b) to stochastic approximations.
\begin{enumerate}
\item Generate $S$ parameters and datasets $(\btheta^{\star s}, \bY^{\star s})_{s = 1,\dots,S}$ according to the Bayesian model $\pi(\btheta^{\star s})\ell(\bY^{\star s} | \btheta^{\star s})$.
\item For each $s = 1\dots S$, from dataset $\bY^{\star s}$, 
\begin{enumerate}
\item derive the deterministic approximation of the posterior $p_{\bY^{\star s}}$ using $\Mcal$,
\item get a sample $(\btheta^{s}_m)_{m = 1\dots M}$ from $\Mcal$.
\end{enumerate}
\item Choose a real-valued function of the parameter $\Phi(\btheta)$ and compute 
\begin{enumerate}
\item $ U(\btheta^{\star s}, \bY^{\star s},\Phi,p_{\bY^{\star s}})$ using \eqref{eq:crit-theo},
\item $U_M(\btheta^{\star s},\bY^{\star s},\Phi, (\btheta^{s}_m)_{m = 1\dots M})$ using \eqref{eq:crit-emp}.
\end{enumerate}
\item Compare the empirical distribution 
\begin{enumerate}
\item of $U(\btheta^{\star s}, \bY^{\star s},\Phi,p_{\bY^{\star s}}))_{s=1\dots S}$  to the uniform distribution on $[0, 1]$,
\item of $\left(U_M(\btheta^{\star s},\bY^{\star s},\Phi, (\btheta^{s}_m)_{m = 1\dots M}\right)_{s=1\dots S}$ to the uniform distribution on $\{1, \dots, \frac{1}{M}\}$. 
\end{enumerate}
\end{enumerate}
\vspace{-1em}
\noindent\rule{\columnwidth}{0.4pt}

\vspace{+0.5em}

The comparison can be performed through graphical tools --~for instance the empirical Cumulative Distribution Function (cdf)~-- or through a statistical test such as the discrete goodness of fit test, discrete version of the Kolmogorov-Smirnov  (KS) test. If we reject the hypothesis $$
H_0 = \left\{U_M(\btheta^{\star},\bY^{\star  },\Phi, (\btheta_m)_{m = 1\dots M}) \sim \mathcal{U}_{\left\{0,\frac{1}{M}, \frac{2}{M},\dots, 1\right\}} \right\}, 
$$
we conclude that the obtained sample is not distributed from the true posterior distribution. 

In our case, several methods $\Mcal$ will be considered: the variational Bayes approximation (VB), the \SBS starting from VB and \SBS starting from a symmetrized version of VB.

\paragraph{Simulation design.} We set $n = 100$ individuals, $p = 10$ observations by individual and $g = 2$ groups in the mixture. Referring to equation \eqref{eq:LCAprior}, the hyper-parameters are set to 
 $\delta = a = b = 2$. $S = 500$ datasets $\bY^{\star s}$ and are simulated. For each simulated dataset, we run the VB algorithm implemented in the BayesLCA package \citep{White13}. The sampling algorithms \SBS starting from $\pt^\mathit{VB}_{\bY^{\star s} }$ and $\pt_{\bY^{\star s} }^\mathit{VB.Sym}$ respectively are implemented with $M = 5000$, $\tau_1 = 0.9$ and $ \tau_2 = 0.9$. The $K_h$ kernel is standard Gibbs algorithm of length $B = 5$.

\subsubsection{Results}

\begin{figure}[ht!]
 \begin{center}
   \includegraphics[width = 0.475\textwidth, height = 0.375\textheight]{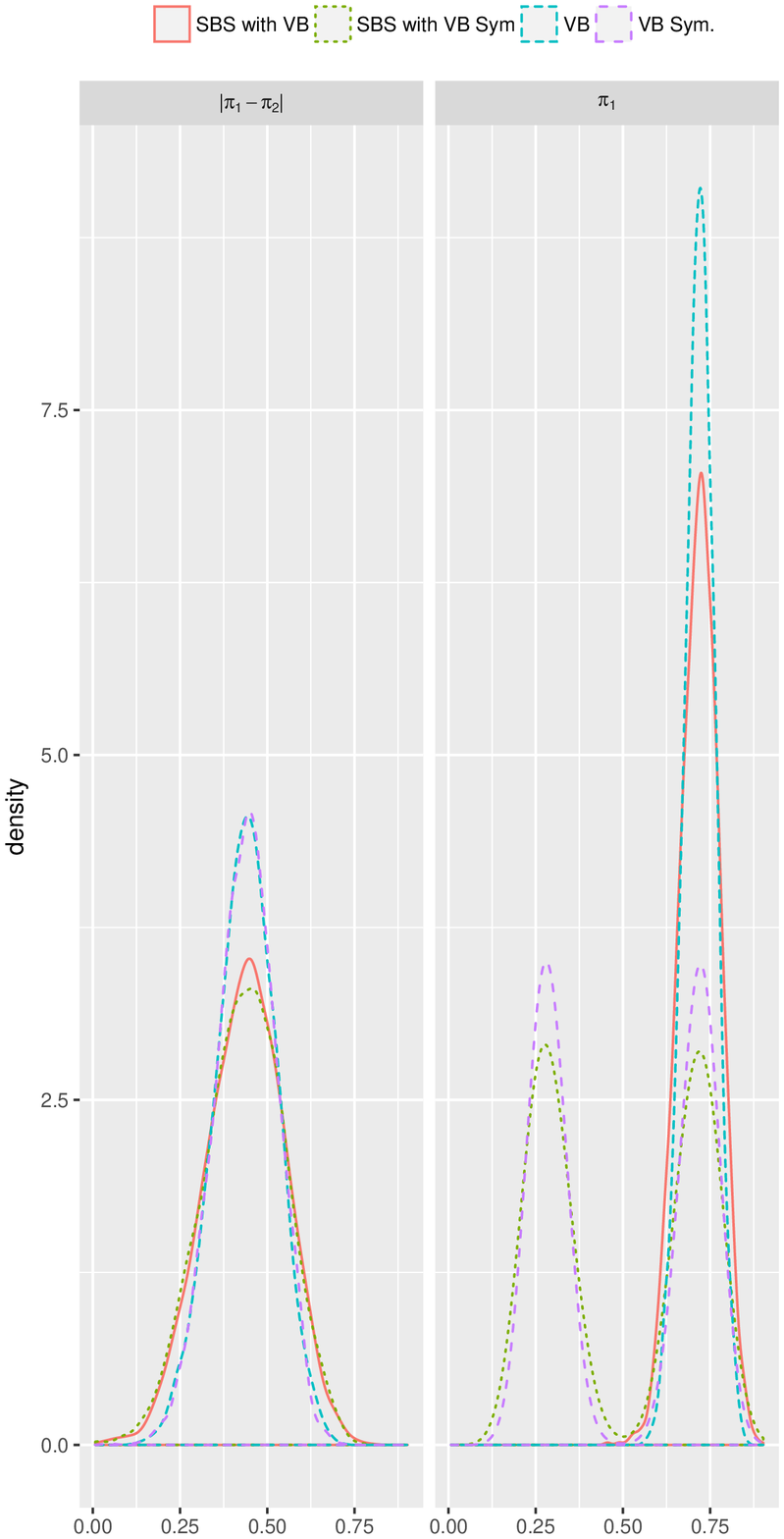} 
  \includegraphics[width = 0.475\textwidth, height = 0.375\textheight]{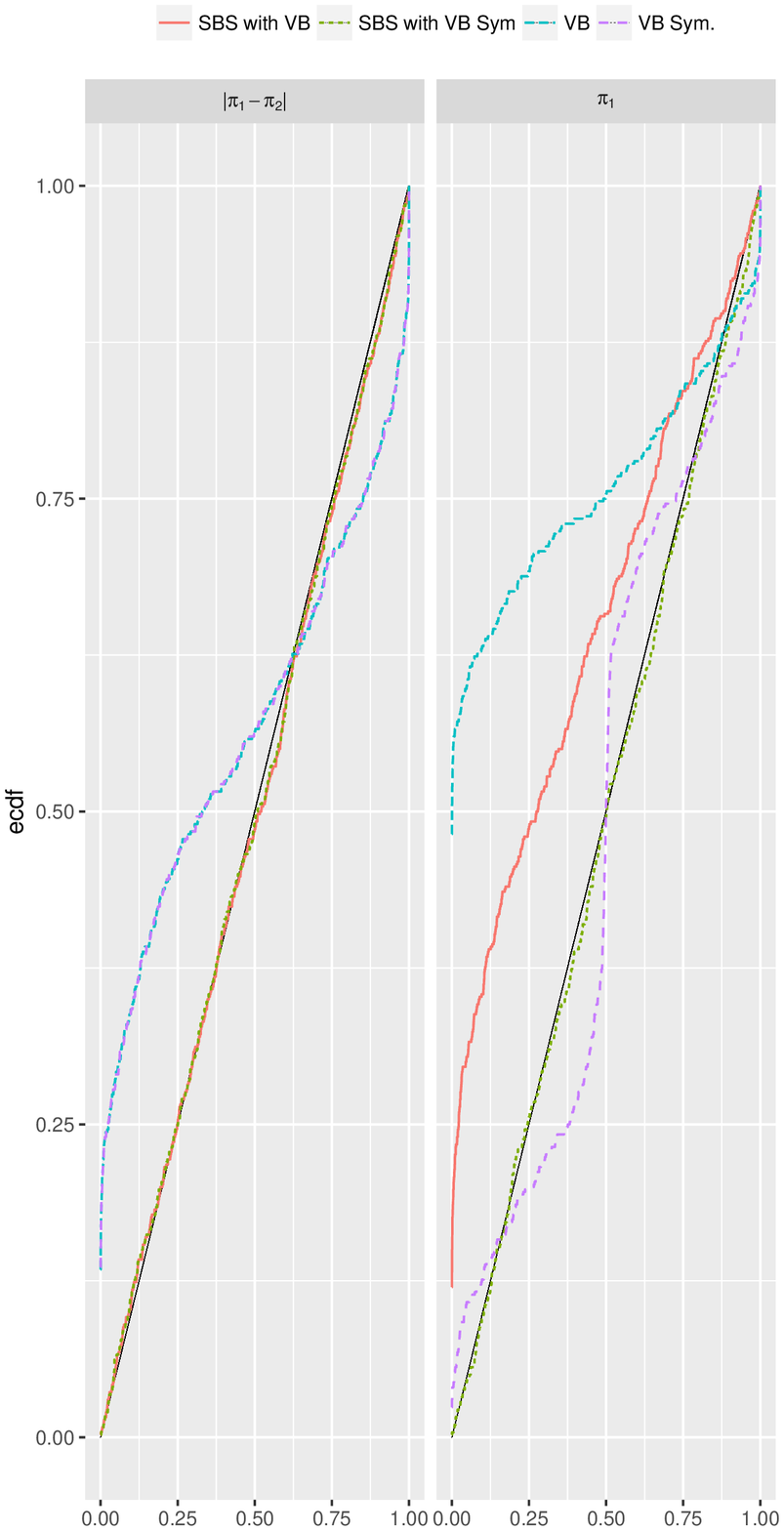} 
 \end{center}
 \caption{\emph{Simulation results for the LCA model.}   Left panel:  posterior distributions for $|\pi_1-\pi_2|$ (left) and $\pi_1$ (right). Right panel:  cumulative distribution function for $U_M(\btheta^{\star s}, \bY^{\star s},\Phi_r,q_{\bY^{\star s}})$The results for $\pt_{\bY^{\star s} }$ and $\pt^ {Sym}_{\bY^{\star s} }$ are in blue and purple respectively. The results obtained by the \SBS algorithm starting from $\pt_{\bY^{\star s} }$ and $\pt^{\mathit{Sym}}_{\bY^{\star s} }$ are in red and green respectively. }
 \label{Fig:post_pi1}
\end{figure}

On Figure \ref{Fig:post_pi1} (left panel), we plot the estimated posterior density (for a arbitrarily chosen dataset) obtained from the Variational approximation, its symmetrized version, the \SBS starting from with VB and the \SBS stating from with VB symmetrized. As expected, the variational approximation underestimates the posterior variance. \SBS succeeds in inflating this posterior variance.  This phenomenon can be observed on $ {\Phi_1(\theta) =}|\pi_1-\pi_2|$ (left) and $ \Phi_2(\theta) = \pi_1$. About $\pi_1$, we observe that using only the classical variational Bayes estimator --highly concentrated on the MAP-- we are enable to explore the other modes. However, if forcing the exploration of the different modes, we are able to observe the full posterior distribution, charging all the modes.

\vs

\noindent On Figure \ref{Fig:post_pi1} (right panel), we plot the empirical cdf's (ecdf) for $U_M$. The results for $\Phi_1$ ($\Phi_2$ respectively) are on the left (right respectively). The ecdf for $\left(U(\btheta^{\star s}, \bY^{\star s},\Phi_l, \pt^{\mathit{VB}}_{\bY^{\star s}})\right)_{s = 1\dots S}$ is in blue and the ecdf for $\left(U(\btheta^{\star s}, \bY^{\star s},\Phi_l, \pt^{\mathit{VB Sym}}_{\bY^{\star s}})\right)_{s = 1\dots S}$ is in purple. The ecdf obtained from $(\btheta^{s}_m)_{m = 1\dots M}$ and $(\btheta^{s, \mathit{Sym}}_m)_{m = 1\dots M}$ sampled by algorithm \SBS starting from $\pt^\mathit{VB}_{\bY^{\star s} }$ ( respectively $\pt_{\bY^{\star s} }^\mathit{VB.Sym}$) are plotted in red (green respectively).

\noindent The phenomena observed before on a single dataset are confirmed here on the $500$ datasets. On $|\pi_1-\pi_2|$,  which is insensitive to label switching,  the non-symmetrized and symmetrized versions of the algorithms give equivalent results (curves red/green, and blue/purple can not be distinguished), which is not the case on $\pi_1$  (which is actually non-identifiable because of label switching). 
On $|\pi_1-\pi_2|$, $\pt^\mathit{VB}_{\bY^{\star s} }$ and $\pt_{\bY^{\star s} }^\mathit{VB.Sym}$ are different from the true posterior distribution but our two algorithms starting respectively from $\pt^\mathit{VB}_{\bY^{\star s} }$ and $\pt_{\bY^{\star s} }^\mathit{VB.Sym}$ supply a sample from the true posterior distribution. 

\noindent On $\pi_1$, we observe that starting \SBS from $\pt^\mathit{VB}_{\bY^{\star s} }$ clearly leads to the wrong posterior. The equality of the true posterior distribution with the one obtained via \SBS starting from VB.Sym    is not rejected.  

\begin{table}
\centering
\begin{tabular}{lcc}
\hline
& $|\pi_1-\pi_2|$ & $\pi_1$\\
\hline \\ VB & $4.497e^{-06}$ & $<2.2e^{-16}$  \\
 Symmetrized VB & $1.563e^{-05}$ & $1.431e^{-09}$\\
 \SBS with VB &   $0.596$ & $<2.2e^{-16}$  \\
 \SBS with Symmetrized VB&  $0.567$ &    $0.903$ \\
\hline
\end{tabular}
\caption{Simulation results for the LCA model: p-values for the  KS test of Property \ref{lem:criterion}.}
\end{table}


\subsection{Stochastic block models with covariates}\label{subsec:SBMreg}

\newcommand{\gtrue}{g_*}

\paragraph{Model.}
As a last example, we consider the combination of stochastic block-model \citep[SBM]{NoS01} and logistic regression (shortened as '\SBMreg' in the sequel) considered in \cite{LRO15}. This model aims at deciphering some residual structure in an observed network once accounted for the effect of some edge covariates. The model is as follows. Consider a set of $n$ nodes;  for each pair ($1 \leq i < j \leq n$) of nodes, we observe a $p$-dimensional covariates vector $\bx_{ij}$. Likewise in SBM, we further assume that each node belongs to one among $g$ groups and we denote $Z_i$ the (unobserved) group where node $i$ is affected; $\bpi = (\pi_k)_k$ denotes the vector of group proportions. The model states that the edges of the observed binary undirected network $\bY = (Y_{ij})$ are drawn independently conditionally on the set of latent variables $\bZ = (Z_i)$ as Bernoulli variables:
$$
(Y_{ij} | Z_i, Z_j, \balpha, \bbeta) \sim \Bcal(p_{ij}), 
\quad 
\logit(p_{ij}) = \bx_{ij}^\intercal \bbeta + \alpha_{Z_i, Z_j}
$$
where $\balpha = (\alpha_{kl})$ stands for the matrix of between-group effects (analogous to the between-group connection probabilities from SBM, in logit scale) and $\bbeta = (\beta_\ell)_{\ell =1,\dots, p}$ for the vector of regression coefficients. As for the priors, $\pi$  has a Dirichlet distribution, both $\alpha$ and $\beta$ are Gaussian. When considering model selection or averaging, the number of groups $g$ is supposed to be uniformly distributed among $\{1, \dots, g_{\max}\}$. 

\paragraph{Bayesian model averaging (BMA).} BMA \citep{HMR99} is a general principle, which consists in combining the results obtained with several models, rather than to choose the 'best' one. Among other interests, it allows to account for model uncertainty. We apply this principle to a regression parameter $\beta_\ell$. While model selection consists in choosing $g$ as $\widehat{g} = \arg\max_g p(g | \bY)$ and considering the posterior $p(\beta_\ell | \bY, g = \widehat{g})$, BMA directly considers the unconditional posterior
$$
p(\beta_\ell | \bY) = \sum_g p(g | \bY) p(\beta_\ell | \bY, g).
$$
In terms of moments,  it results in $\Esp(\beta_\ell | \bY) = \sum_g p(g | \bY) \Esp(\beta_\ell | \bY, g)$ and
$\Var(\beta_\ell | \bY) = \Var_{\text{within}}(\beta_\ell | \bY) + \Var_{\text{between}}(\beta_\ell | \bY)$
where $\Var_{\text{within}}$ measures the mean variance of the parameter conditionally on $g$ and $\Var_{\text{between}}$ is  the variance of the parameter due to model uncertainty:
\begin{eqnarray*}
\Var_{\text{within}}(\beta_\ell | \bY) & = & \sum_g p(g | \bY) \Var(\beta_\ell | \bY, g), \\
\Var_{\text{between}}(\beta_\ell | \bY) & = & \sum_g p(g | \bY) \left(\Esp(\beta_\ell | \bY, g) - \Esp(\beta_\ell | \bY)\right)^2.
\end{eqnarray*}

\paragraph{VB approximation.} As illustrated in Subsection \ref{subsec:SBMreg}, the VB approximate posterior is quite accurate for logistic regression and \cite{GDR12} also proved its empirical accuracy for SBM. A first goal of this simulation study is to check if this accuracy still holds when the two models are combined into the \SBMreg model. To this aim, we focus on the posterior distribution of the regression parameters. Secondly, we want to check the accuracy of the VB posterior distribution of the number of groups, that can be used either to assess goodness-of-fit or for model averaging (\cite{LRO15})

\paragraph{Simulation design.} 
We simulate networks with $n \in \{20, 50\}$ nodes according to an \SBMreg model with $\gtrue \in \{1, 2\}$ groups and $p = 3$ covariates.  To apply  Property \ref{lem:criterion}, the parameters are sampled from the prior distribution.  $S = 100$ replicates are simulated for each configuration and for each of them , the  \SBMreg models  with $g \in \{1, \dots, g_ {\max} = 5\}$ were fitted with the VB algorithm described in \cite{LRO15}. The \SBS algorithm is then run on each dataset.

\paragraph{Results for parameter estimation.}

\begin{figure}[ht!]
 \begin{center}
  \begin{tabular}{c|c}
      \includegraphics[width=.4\columnwidth, height=.4\textwidth]{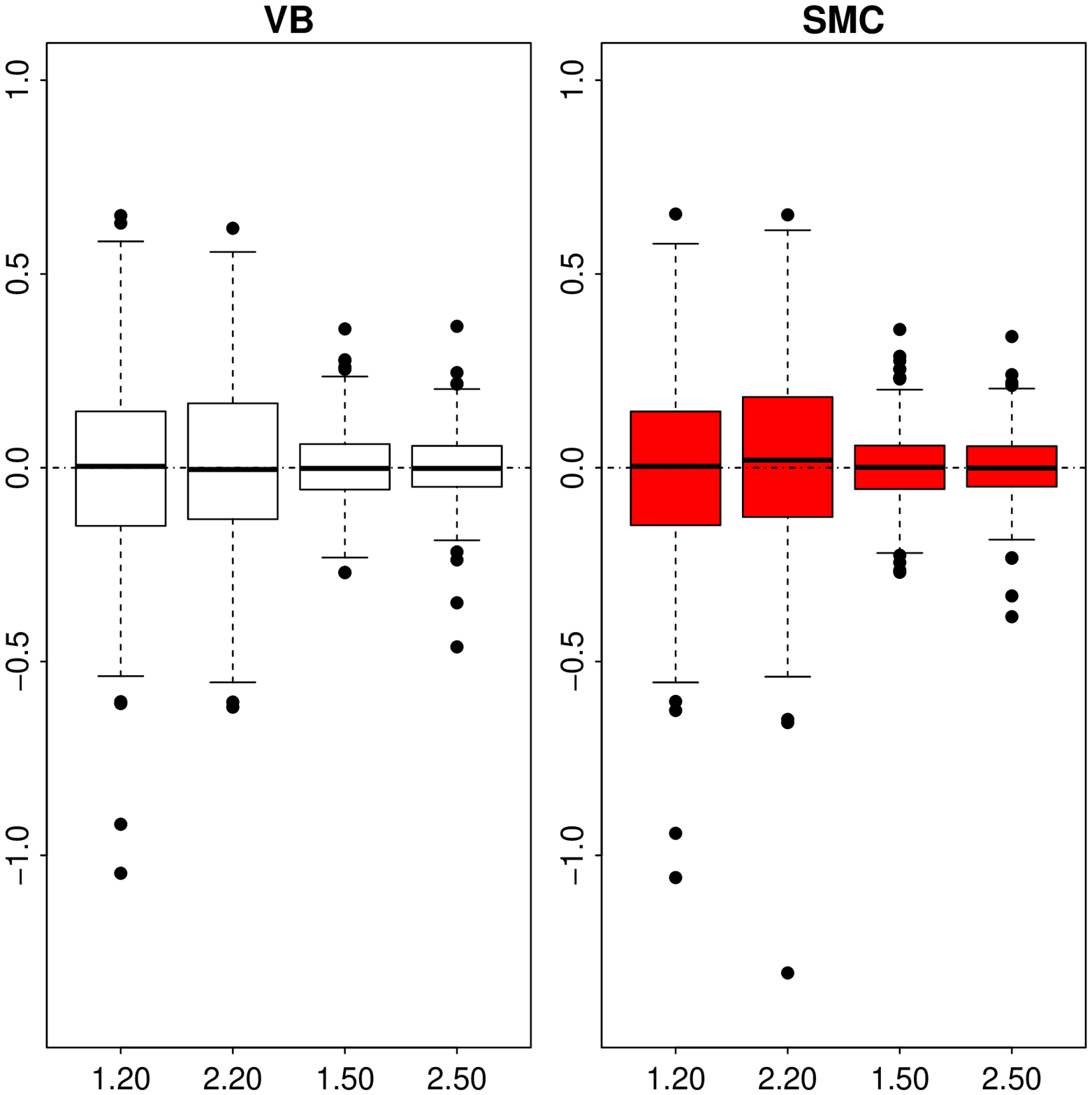} 
   & 
   \includegraphics[width=.4\columnwidth, height=.4\textwidth]{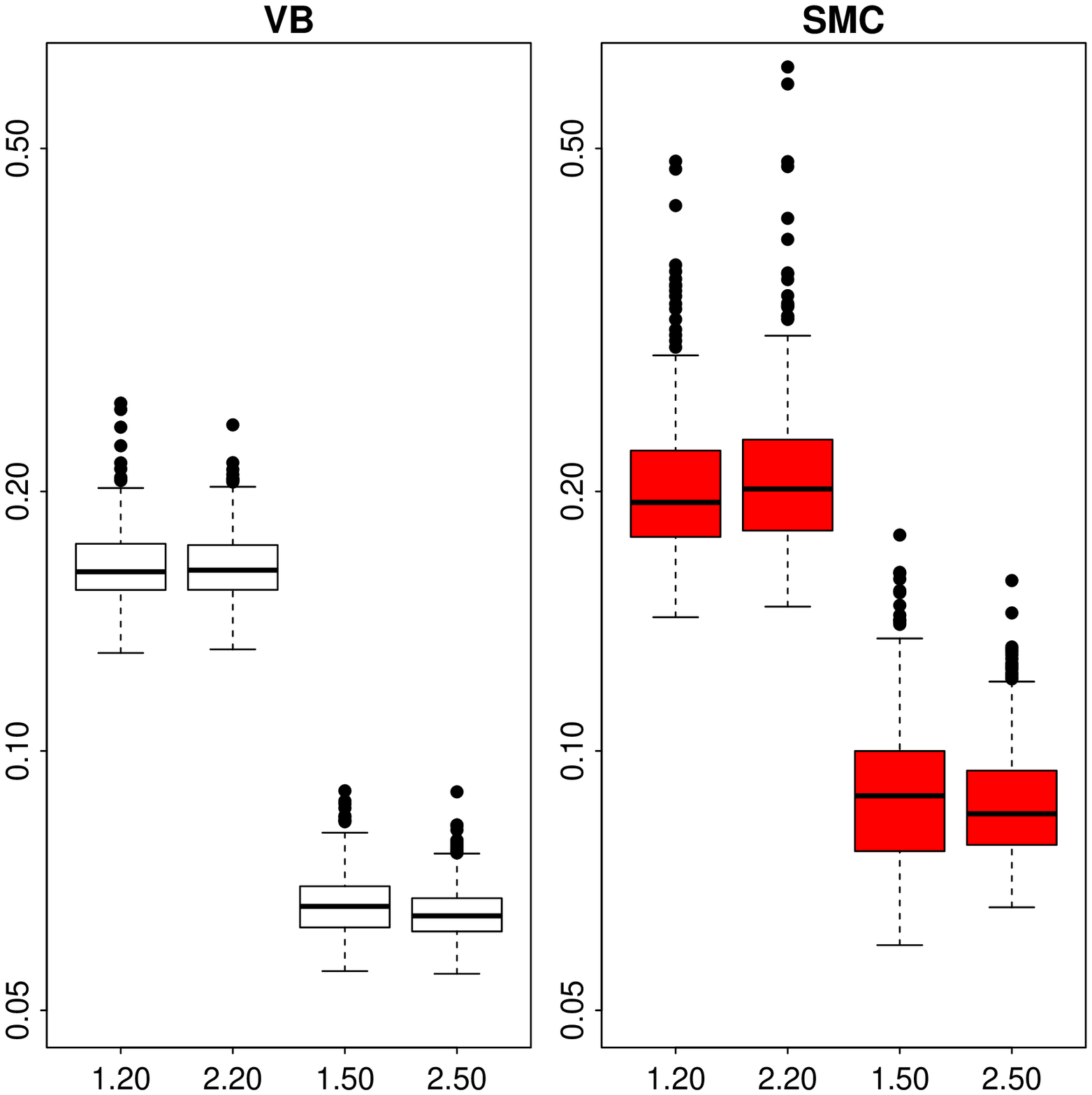} \\
   \hline
   \includegraphics[width=.4\columnwidth, height=.4\textwidth]{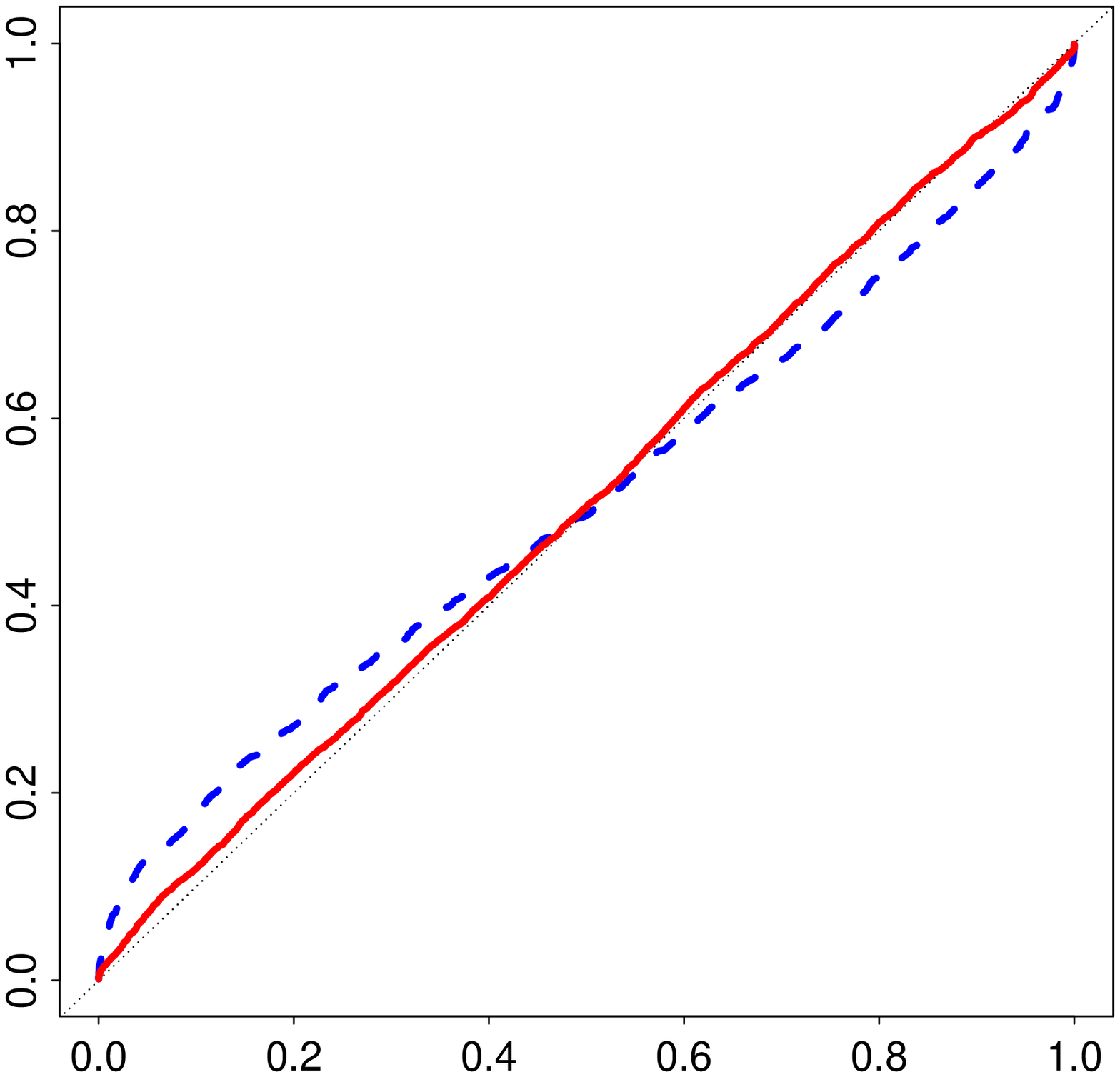} 
   & 
   \includegraphics[width=.4\columnwidth, height=.4\textwidth]{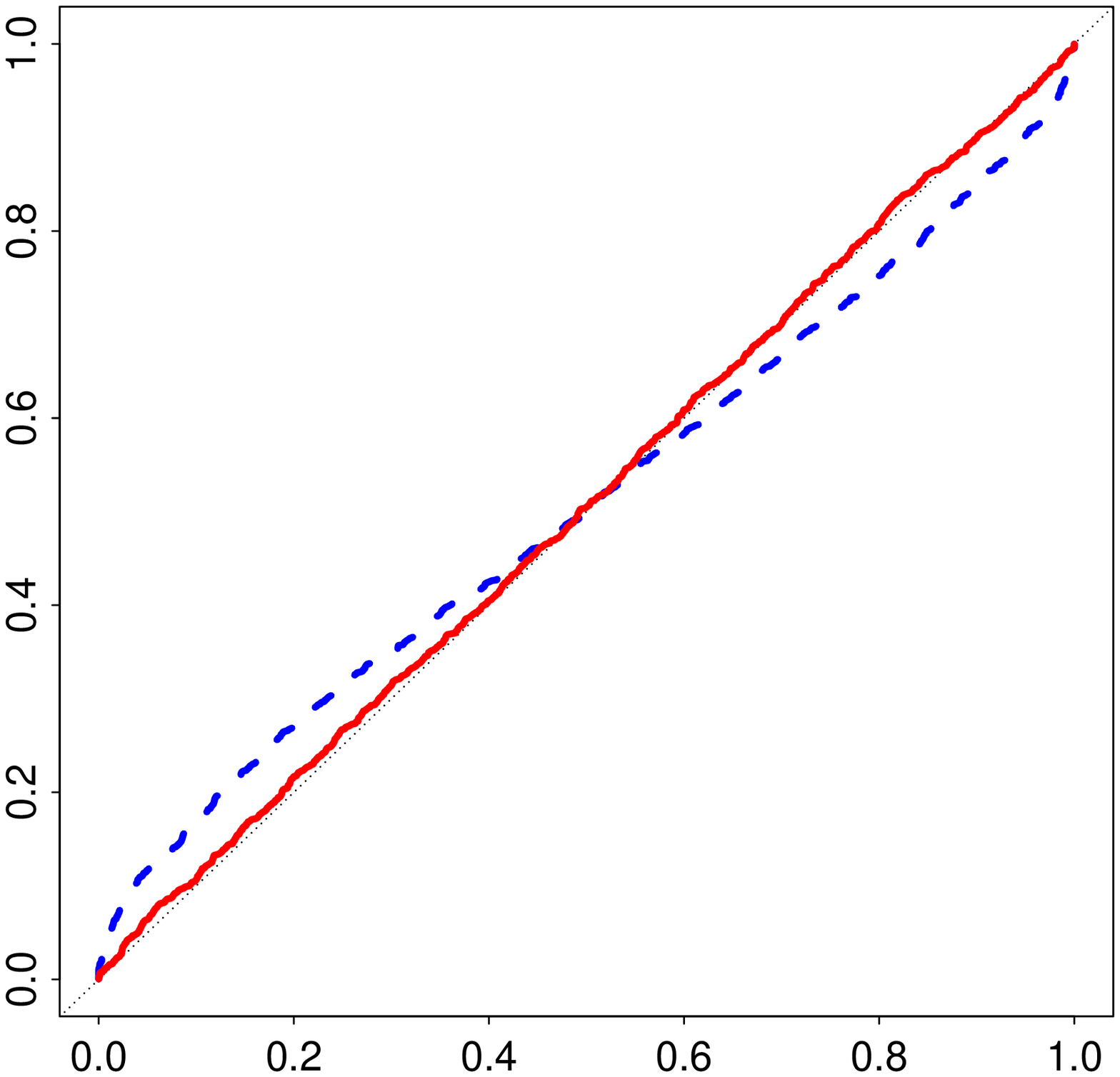} 
  \end{tabular}
 \end{center}
 \caption{Simulation results for the \SBMreg model: VB (white) and \SBS (red) posterior of the regression coefficients $\bbeta = (\beta_\ell)$. Top: posterior mean (left), posterior standard deviation (right); $x$-axis label: $\gtrue . n$ (e.g. '1.20' means $\gtrue = 1$, $n= 20$). 
Bottom:
graphical check of Property \ref{lem:criterion} for VB  (dashed blue) and 
for \SBS (solid red). Left: $g$ = $\gtrue$, right: with model averaging.
}
\label{Fig:SimulSMBreg-beta} 
\end{figure}

We first consider the posterior distribution of $\beta$ when the number of groups $g$ is known.  On Figure~\ref{Fig:SimulSMBreg-beta}, we plot on the left the boxplots for the posterior means $(\widehat{\mathbb{E}}^{VB}(\beta_\ell | \bY^s))_{s=1\dots 100}$ and $(\widehat{\mathbb{E}}^{\SBS}(\beta_\ell | \bY^s))_{s=1\dots 100}$.  The boxplot (over the 100 simulated datasets)  of the posterior standard deviations $(\widehat{\sigma}^{VB}(\beta_\ell | \bY^s))_{s=1\dots 100}$ and $(\widehat{\sigma}^{\SBS}(\beta_\ell | \bY^s))_{s=1\dots 100}$ are on the top-right. 
We clearly observe that the posterior means provided by VB and \SBS are both accurate and similar, but the VB's posterior standard deviations (sd) are smaller than \SBS's posterior standard deviations.

To further assess the quality of the posterior distribution provided by VB and \SBS, 
we  checked Property  \ref{lem:criterion}  for  the regression coefficients, which are not subject to label switching. 
We observe that the ecdf of the \SBS  sample is close to uniform, whereas there is a departure for VB.  The p-values resulting from  KS test of Property \ref{lem:criterion} (see Table \ref{Tab:pvalSBMreg}, upper table) lead to the same conclusions. All these observations concur to show that, although the VB approximate posterior distribution is accurate for logistic regression and SBM separately, it is biased for the \SBMreg model, and that the proposed  \SBS is a way to correct it. 
As a consequence of this phenomenon, the empirical level of VB's credibility intervals is    equal to 84.75\%, which is below the nominal level  95\%, whereas \SBS's credibility intervals almost reach the targeted level (93.75\%).

\begin{table}[ht!]
 \begin{center}
  \begin{tabular}{ccccc}
  & \multicolumn{4}{c}{$\widehat{g} = \gtrue$} \\
  \hline
  & \multicolumn{2}{c}{$n = 20$} & \multicolumn{2}{c}{$n = 50$} \\
  & $g = 1$ & $g = 2$ & $g = 1$ & $g = 2$ \\
  \hline
  VB & 0.027 &  0.004 & 0.002 & 0.077  \\
  \SBS & 0.785 & 0.121 & 0.839 & 0.238  \\
  \hline
  $\;$& $\;$&$\;$&$\;$\\
   $\;$& $\;$&$\;$&$\;$\\
  & \multicolumn{4}{c}{BMA} \\ 
  \hline 
 & \multicolumn{2}{c}{$n = 20$} & \multicolumn{2}{c}{$n = 50$} \\
  & $g = 1$ & $g = 2$ & $g = 1$ & $g = 2$\\
   \hline
  VB & 0.017 & 0.003 & 0.002 & 0.079 \\
   \SBS  & 0.740 & 0.277 & 0.778 & 0.312\\
  \end{tabular}
  \caption{Simulation results for the \SBMreg model: p-values   for the KS test of Property \ref{lem:criterion}. \label{Tab:pvalSBMreg}}
 \end{center}
\end{table}

\begin{figure}[ht!]
 \begin{center}
  \begin{tabular}{c|c}
    \includegraphics[width=.4\columnwidth, height=.3\textwidth]{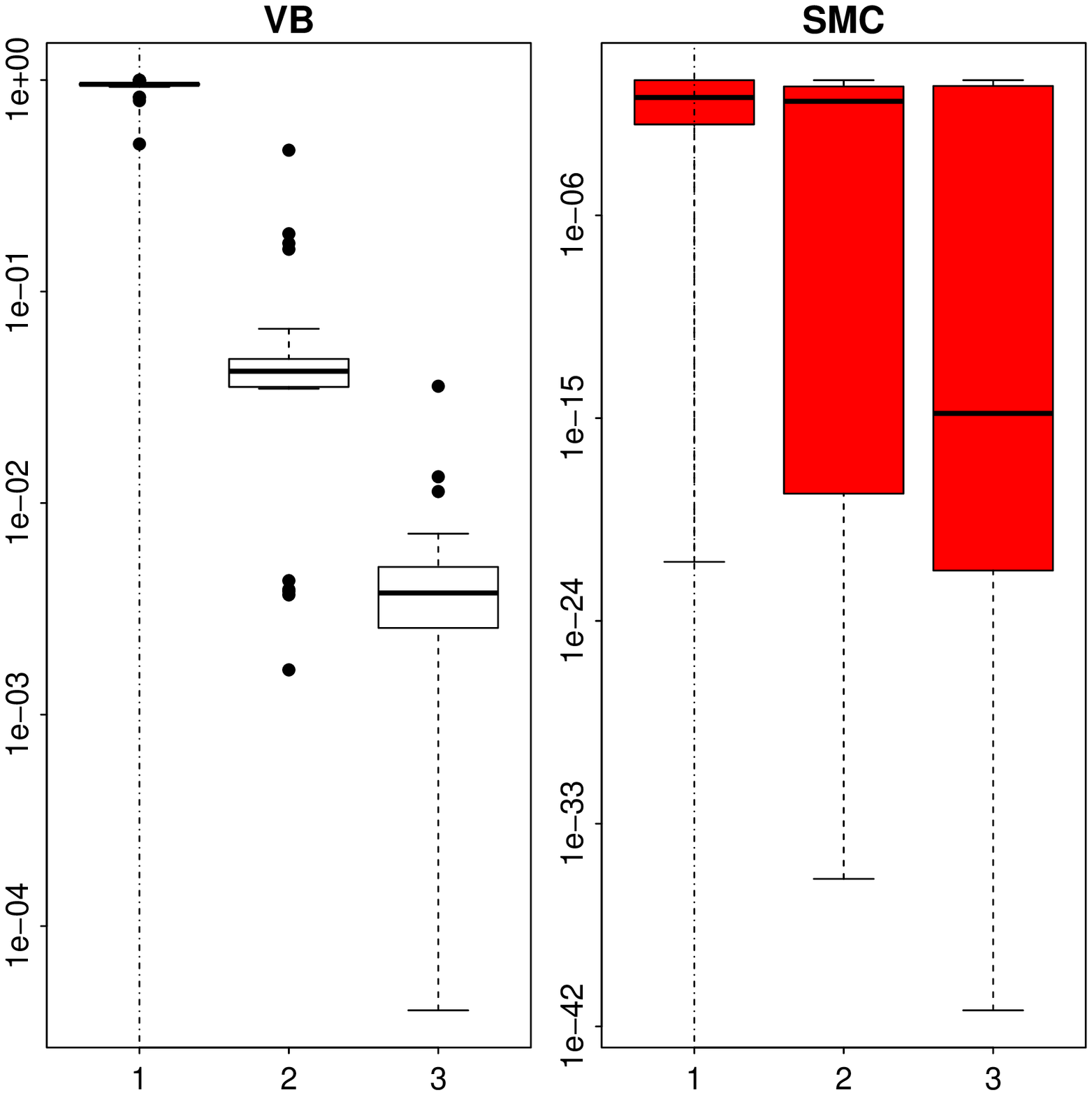} &
    \includegraphics[width=.4\columnwidth, height=.3\textwidth]{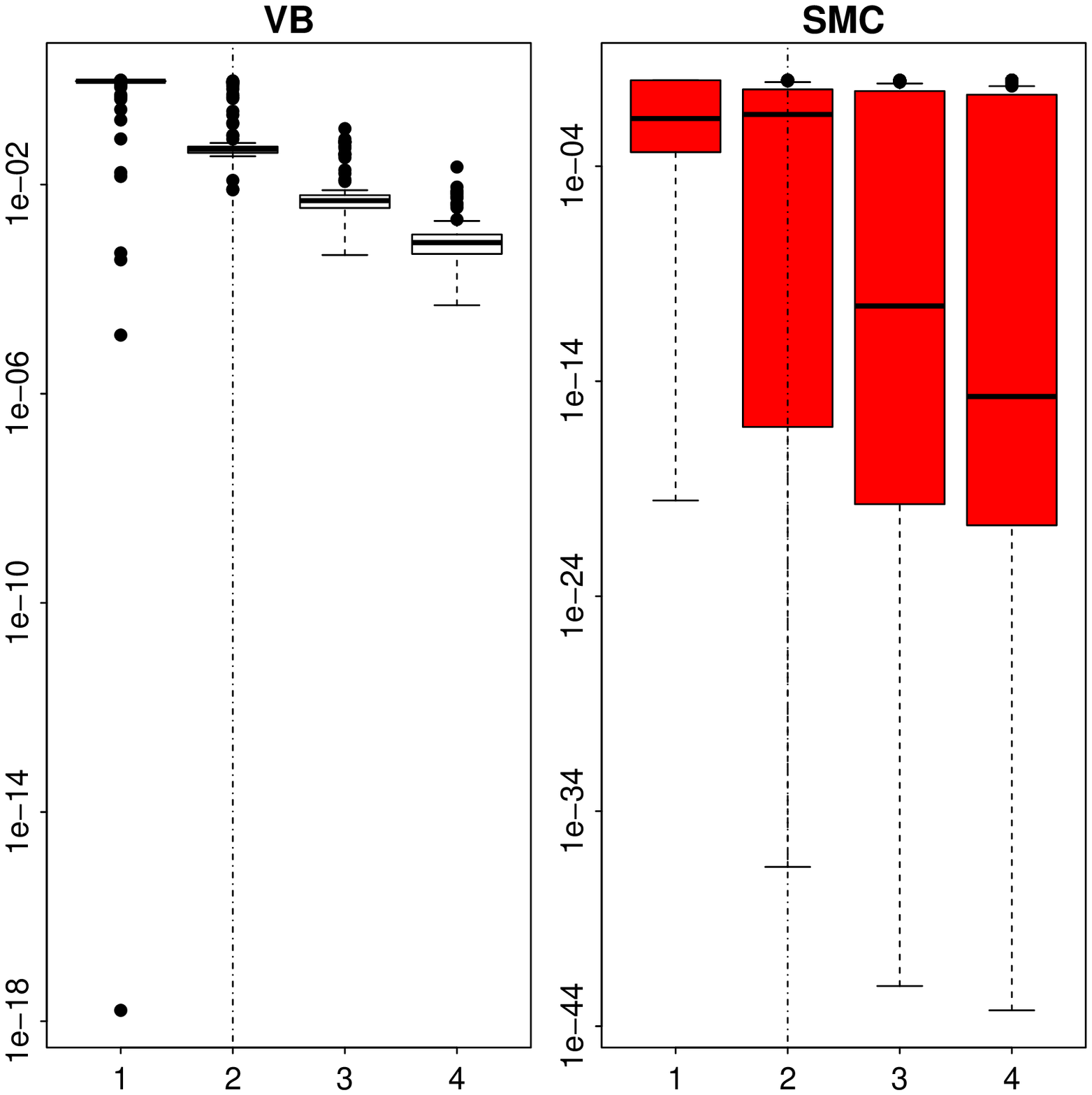} \\ 
    \hline
    \includegraphics[width=.4\columnwidth, height=.3\textwidth]{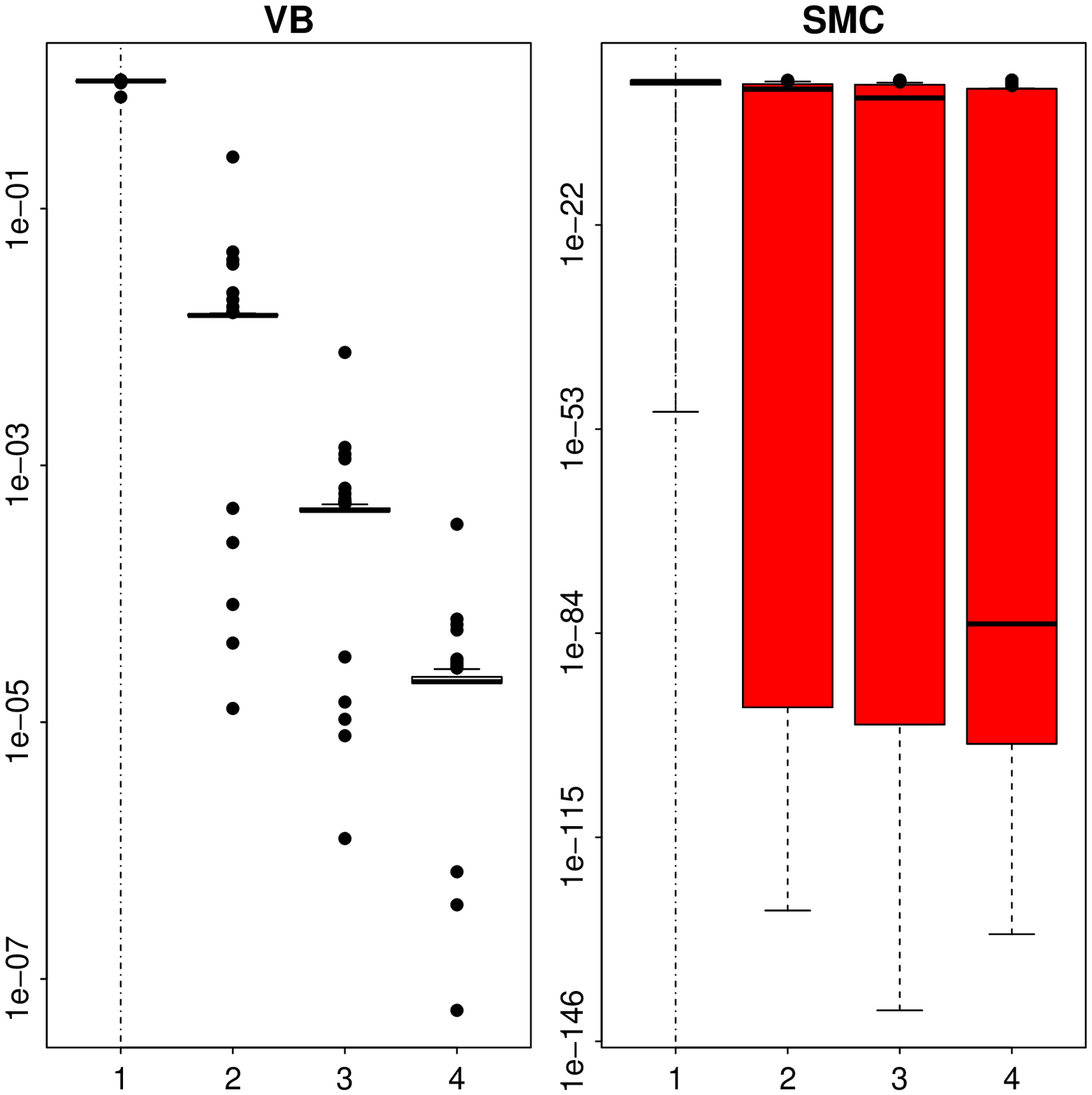} &
    \includegraphics[width=.4\columnwidth, height=.3\textwidth]{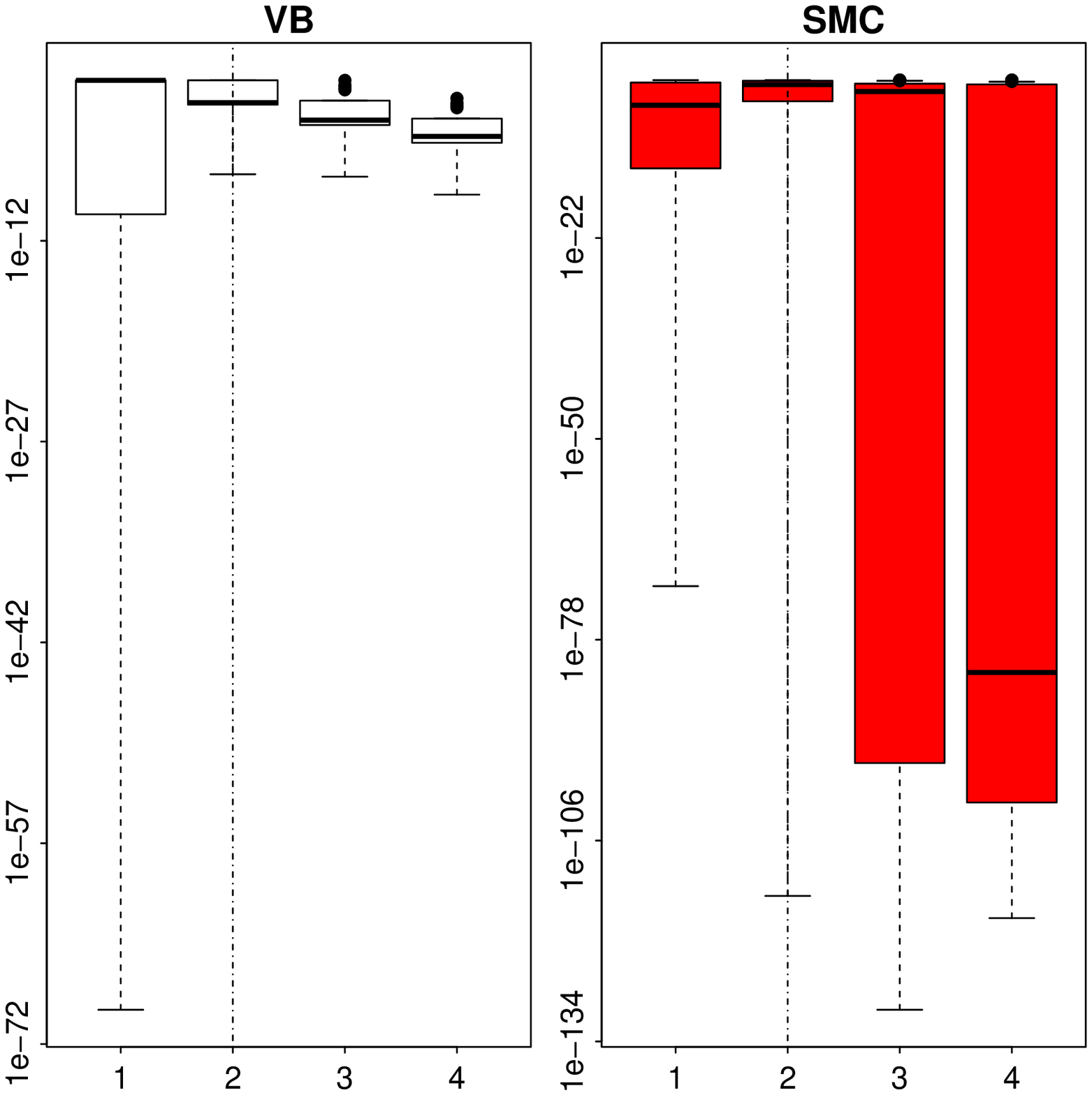} 
  \end{tabular}
 \end{center}
 \caption{Simulation results for the \SBMreg model: box-plots for the posterior probability $p(g|\bY)$ as a function of $g$. Top $n=20$, bottom: $n=50$. Left: $\gtrue = 1$, right: $\gtrue = 2$. 
 \label{Fig:SimulSMBreg-modsel}}
\end{figure}

\paragraph{Results for model selection.}
We now consider the posterior distribution of the number of groups $p(g|\bY)$ and its use for model selection. Figure \ref{Fig:SimulSMBreg-modsel} provides a comparison of the posterior provided by VB and  \SBS. We observe that the VB approximation always results in a more concentrated distribution than \SBS. This behavior can be compared to the under-estimation of the posterior variance of the parameters that we already discussed. To compare the results in terms of model selection we computed the frequency at which the right model is selected (i.e. when $\widehat{g} = \gtrue$) and the mean posterior probability of the $\gtrue$ (see Table \ref{Tab:SimulSMBreg-Pgstar}). We observe that VB performs better than \SBS for both criteria. This parallels \cite{Min05}, who shows that the minimization of the Kullback-Leibler (KL) divergence leads to an accurate estimate of the mode, which is convenient for model selection.

\begin{table}[ht!]
 \begin{center}
  \begin{tabular}{ccccc}
  & \multicolumn{4}{c}{frequency of $\widehat{g} = \gtrue$ (\%)}\\
  \hline
  & \multicolumn{2}{c}{$n=20$} & \multicolumn{2}{c}{$n=50$} \\
  & $\gtrue=1$ & $\gtrue=2$ & $\gtrue=1$ & $\gtrue=2$ \\
  \hline
  VB & 100 & 10 & 100 & 42  \\
  \SBS & 46 & 23 & 60 & 36  \\
   \hline
  $\;$& $\;$&$\;$&$\;$\\
   $\;$& $\;$&$\;$&$\;$\\
   & \multicolumn{4}{c}{mean value of $p(g = \gtrue | \bY)$} \\
     \hline
    & \multicolumn{2}{c}{$n=20$} & \multicolumn{2}{c}{$n=50$}\\
     & $\gtrue=1$ & $\gtrue=2$ & $\gtrue=1$ & $\gtrue=2$ \\
       \hline
     VB & 0.947 & 0.138 & 0.982 & 0.410 \\
  \SBS &0.435& 0.257 & 0.562 & 0.387 \\
  
  \end{tabular}
  \caption{Simulation results for the \SBMreg model: model selection \label{Tab:SimulSMBreg-Pgstar}}
 \end{center}
\end{table}

\noindent Although it does not seem to hamper model selection, the biased estimation of the posterior $p(g|\bY)$ may have undesired consequences when  used for model averaging. To illustrate this point, we simply computed the empirical coverage of credibility intervals for each $\beta_\ell$ after model averaging. The mean coverage across simulation condition and covariate index $\ell$ for VB (85.8\%) is still below the nominal level, whereas this of  \SBS (93.25\%) is close to 95\%. Figure \ref{Fig:SimulSMBreg-beta} (bottom right) also shows that the distribution of the ecdf after model averaging is almost confounded with the uniform for  \SBS, whereas it still displays a significant bias for VB.  The p-values  for the KS test of Property \ref{lem:criterion} (see Table \ref{Tab:pvalSBMreg}, bottom) lead to the same conclusion.

\section{Illustrations  on network datasets} \label{sec:Illust}

\paragraph{Network analysis with \SBMreg.}

To illustrate the use of the proposed sampling algorithm, studied a series of examples analyzed by \cite{LRO15} with an \SBMreg model. The main two purposes of such an analysis is ($i$) to estimate the effect $\bbeta$ of the covariates and ($ii$) to assess the goodness-of-fit of the model based on the covariates. Task ($ii$) is achieved by computing the posterior probability for the SBM part of the model to involve only $g=1$ class, that is $p(g = 1 | \bY)$. A low value of this probability is an indication for a residual structure in the network.

We refer to \cite{LRO15} for the presentation of the data. We considered the datasets (networks) refereed to as Florentine (business), Florentine (marriage), Karate, Tree and Blog. Their respective sizes range from few tens to few hundreds nodes and their densities from 1\% to 50\%. Note that the numerical results presented here for the VB inference slightly differ, as we kept all nodes from each graph whereas \cite{LRO15} removed all isolated nodes.

\begin{figure}[ht!]
 \begin{center}
  \begin{tabular}{cc}
     \includegraphics[width=.4\columnwidth]{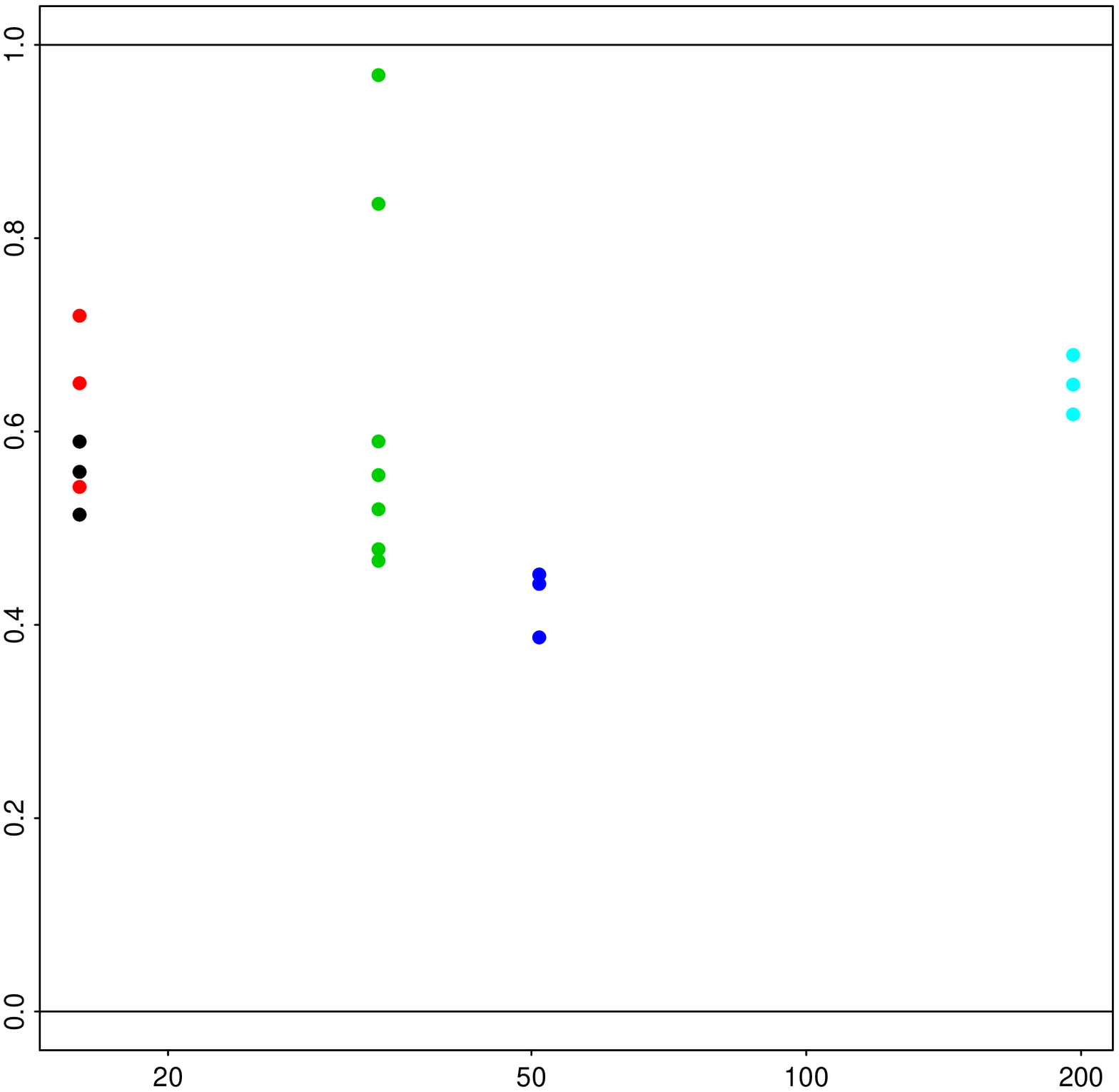} &
   \includegraphics[width=.4\columnwidth]{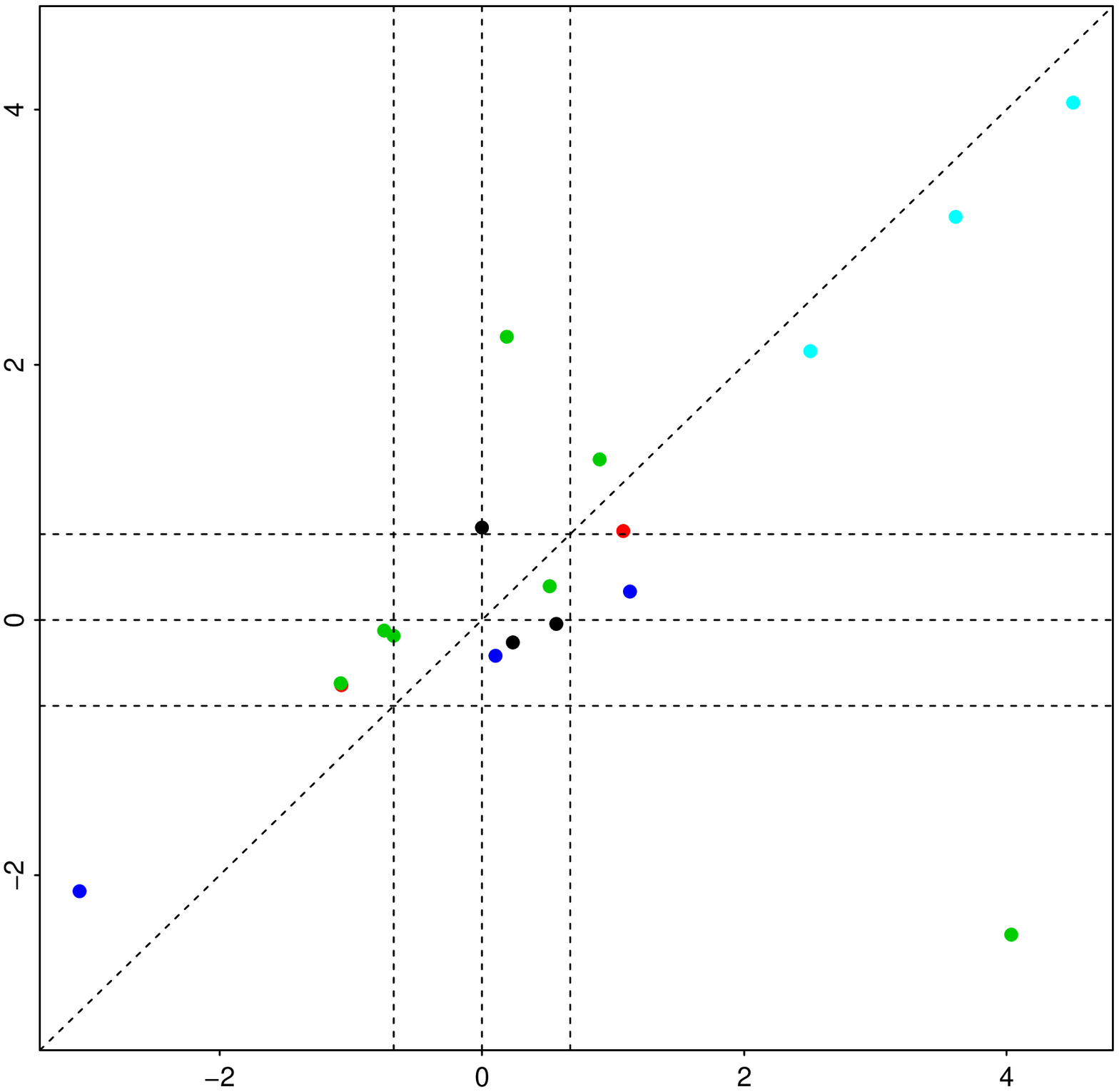}
  \end{tabular}
  \caption{Posterior moments of the regression coefficients for the datasets from \cite{LRO15}. Left: ratio between the VB and \SBS sd as a function of the network size (color = network). Right: ratio between the posterior mean and the posterior sd ($x$: VB, $y$: \SBS, signed-log scale, color: network, horizontal and vertical lines: 2.5\% and 97.5\% $\mathcal{N}(0, 1)$ quantiles) \label{Fig:IllustPostBeta}}
 \end{center}
\end{figure}

 For each of these datasets, we fitted an \SBMreg model with $g = 1 \dots g_{\max}$ groups with a VBEM algorithm to obtain the Gaussian VB approximate distribution for $\papprox^{VB}(\bbeta)$. We also run the proposed \SBS   with $M = 1000$ particles and obtained a weighted sample from $\widehat{p}_{\bY}^{\SBS}(\bbeta)$.
To compare the posterior distributions of $\bbeta$, we adopted the Bayesian model averaging principle described in Section \ref{subsec:SBMreg}. For each dataset $s$ and each covariate $\ell$, we first computed the ratio between the VB and \SBS posterior standard deviation (sd)
$\sqrt{\Var^{VB}(\beta_j) / \Var^{\SBS}(\beta_j)}$. 
For each configuration, we also computed the ratios 
$\widetilde{\bbeta}^{VB}_j / \sqrt{\widetilde{\Sigma}^{VB}_{jj}}$ and $\widetilde{\bbeta}^{\SBS}_j / \sqrt{\widetilde{\Sigma}^{\SBS}_{jj}}$, 
which are typically used to evaluate the effect of the covariates. Figure \ref{Fig:IllustPostBeta}, left, shows that the VB approximation tends to under estimate the posterior variance of the parameter. As expected, Figure \ref{Fig:IllustPostBeta}, right, shows that it yields in over-estimating the significance of the effect of the covariates.

To further illustrate these differences, we studied the posterior distribution of the three regression coefficients used in the 'Tree' example. The regression coefficient are associated with the genetic, geographic and taxonomic distance between the tree species. The results given in Table \ref{Tab:IllustPostBeta} indicate again that the posterior sd provided by \SBS are all larger than these resulting from VB. We observe that, because of the high concentration of the posterior distribution of $g$, VB strongly under-estimates the variance due to model uncertainty. A result is that the \SBS posterior standard deviations are almost twice larger than the VB posterior standard deviations. As a consequence, the significance ratio is reduced, and the influence of both the genetic and the geographic distance ($\beta_1$ and $\beta_2$) turn out to be more questionable according to \SBS than to VB.

\begin{table}[ht!]
 \begin{center}
  \begin{tabular}{lccc}
  & \multicolumn{3}{c}{$\widetilde{p}^{VB}_{\bY}$}\\
  \hline
  & $\beta_1$ & $\beta_2$ & $\beta_3$  \\ \hline
post. mean & 4.62e-05 & 0.23 & -0.9 \\ 
post. within var. & 2.24e-10 & 0.0432 & 0.00175  \\ 
post. between var. & 5.55e-17 & 1.18e-06 & 2.42e-07 \\ 
posterior sd & 1.5e-05 & 0.208 & 0.0418 \\ 
ratio & 3.09 & 1.11 & -21.5 \\
  $\;$& $\;$&$\;$&$\;$\\
   $\;$& $\;$&$\;$&$\;$\\
   & \multicolumn{3}{c}{$\widehat{p}^{\SBS}_{\bY}$} \\
   \hline
  &  $\beta_1$ & $\beta_2$ & $\beta_3$ \\ \hline
post. mean  & 4.13e-05 & 0.355 & -0.906 \\ 
post. within var.& 1.09e-09 & 0.219 & 0.00889 \\ 
post. between var.& 3.99e-12 & 0.0019 & 0.00281 \\ 
posterior sd & 3.31e-05 & 0.47 & 0.108 \\ 
ratio& 1.25 & 0.755 & -8.38

\end{tabular}
  \caption{Posterior moments of the regression coefficients. \label{Tab:IllustPostBeta}}
 \end{center}
\end{table}

For the goodness-of-fit study, we compared the values of $\widetilde{p}_{\bY}^{VB}(1)$ and $\widehat{p}_{\bY}^{\SBS}(1)$ (Table \ref{Tab:IllustPostPrM1}). Except in the most uncertain case (Karate), the posterior probabilities are similar and lead to the same conclusion about the existence of a residual structure in the network.

\begin{table}[ht!]
 \begin{center}
  \begin{tabular}{lcc}
     &$\widetilde{p}_{\bY}^{VB}(1)$  & $\widehat{p}_{\bY}^{\SBS}(1)$\\
     \hline
 Marriage & $9.54 \; 10^{-1}$ & $1.00$  \\
   Business &  $7.04 \; 10^{-1}$& $1.00$\\
   Karate&  $2.56 \; 10^{-1}$ &  $7.07 \; 10^{-3}$\\ 
   Tree& $4.83 \; 10^{-153}$  & $1.06 \; 10^{-161}$   \\
    Blog & $8.63 \; 10^{-174}$&  $4.04 \; 10^{-290}$ \\  
  \end{tabular}
  \caption{Posterior probability for the \SBMreg model with only one class. \label{Tab:IllustPostPrM1}}
 \end{center}
\end{table}
\section{Discussion and perspectives} \label{sec:Discuss}

 In this paper, we presented a simple strategy to combine the strength of deterministic approximations of the posterior distribution with sequential Monte Carlo samplers. We illustrated the efficiency  of our approach and its  robustness  with respect to the deterministic approximation on a large simulation study. Its application on network datasets stresses the fact that the well-known underestimation of the posterior variance by the variational approximation can be easily corrected, sometimes leading to different statistical conclusions.  Besides,  if dependencies between parameters have been neglected in the deterministic posterior approximation, they will be recovered by the sequential sampling.

Our approach is not restricted to the case where a standard deterministic posterior approximation can be derived (such as Variational Bayes, Laplace  or Expectation Propagation estimate). Any point estimate can be used to design a rough posterior (using a Gaussian or a log-Gaussian seems to be the simplest solution) and serves as an accelerator of the sampling sequence. This strategy is different from an empirical Bayes strategy, the point estimate being only used to explore more efficiently the posterior distribution and not to elicit a prior distribution. The method is not as sensible as standard Importance Sampling to an eventual under-evaluation of the approximate posterior variance : even with a too narrow approximation of the posterior distribution, the algorithm is able to get back to the true posterior variance.

SMC directly supplies a final population of particles arising from the true posterior distribution, as opposed to MCMC strategies, whose convergence is difficult to assess. The proposed SBS algorithm is adaptive in the sense that the sequence $\papprox(\btheta)^ {1-\rhoh}(p( \btheta | \bY)) ^{\rho_h}$ is determined on the fly in an automatic way. Furthermore, the algorithm path (summarized by the sequence $\rho_h$) is an indicator of the quality of the deterministic posterior distribution used to initiate the bridge sampling. 

A natural extension of the present work is its adaptation to Approximate Bayesian Computation (ABC) context for models with no explicit likelihood, following \cite{DelMoral2012}. The difficulty will arise from the specification of the distributions sequence.

\section*{Acknowledgements}
The authors are thankful to Nicolas Chopin for fruitful discussions.

\bibliographystyle{spbasic}      
\bibliography{biblio}

\end{document}